\documentstyle[pre,aps,epsf]{revtex}
\input{psfig.tex}
\begin{document}


\title{Real Arnold complexity versus real topological entropy
for  birational transformations}

\author{N. Abarenkova}
\address{Centre de Recherches sur les Tr\`es Basses Temp\'eratures,
B.P. 166, F-38042 Grenoble, France\\
Theoretical Physics Department, Sankt Petersburg State University,
Ulyanovskaya 1, 198904 Sankt Petersburg, Russia}

\author{J.-Ch. Angl\`es d'Auriac\footnote{e-mail
: dauriac@crtbt.polycnrs-gre.fr}}
\address{Centre de Recherches sur les Tr\`es Basses Temp\'eratures,
B.P. 166, F-38042 Grenoble, France}

\author{S. Boukraa}
\address{ LPTHE, Tour 16, 1er \'etage, 4 Place Jussieu, 
75252 Paris Cedex, France\\
 Institut d'A\'eronautique, Universit\'e de Blida, BP 270,
Blida, Algeria}

\author{S. Hassani\footnote{e-mail
: hassani@ist.cerist.dz}}
\address{CDTN, Boulevard F.Fanon, 16000 Alger, Algeria}

\author{J.-M. Maillard\footnote{e-mail
: maillard@lpthe.jussieu.fr}}
\address{ LPTHE, Tour 16, 1er \'etage, 4 Place Jussieu, 
75252 Paris Cedex, France}


\maketitle

\begin{abstract}
We consider  a family of 
birational transformations of two variables,
depending on one parameter, for which
 simple rational expressions with integer coefficients,
 for the exact expression of the 
dynamical zeta function, have been conjectured. 
together with an equality between the (asymptotic
of the) Arnold complexity
and the (exponential of the) topological entropy. This
 identification takes place
for the birational mapping seen as a mapping bearing on two complex
variables (acting in a complex projective space).
We revisit this
 identification  between these
two quite ``universal complexities''
 by considering now the 
 mapping as a mapping bearing on two 
{\em real} variables. The definitions of the two previous 
``topological'' complexities (Arnold complexity
and topological entropy) are modified according to this 
real-variables point of view. Most of the ``universality'' 
is lost. However, the results presented here 
are, again, in agreement with an identification
between the  (asymptotic
of some) ``real Arnold complexity'' and the  (exponential of some) 
``real topological entropy''. A detailed analysis of
this  ``real Arnold complexity'' as a function of the 
parameter of this family of birational transformations of two variables
is given. One can also slightly modify the definition
of the dynamical zeta function, introducing
a ``real dynamical zeta function'' associated with the counting
of the real cycles only. Similarly 
one can also introduce some ``real Arnold complexity''
generating functions. We show that several of these 
two ``real'' generating functions seem to have 
the same singularities. Furthermore we actually 
conjecture several simple rational expressions 
for them, yielding again 
algebraic values for the (exponential of the) 
``real topological entropy''. In particular, when
 the  parameter of our family
of birational transformations
 becomes large, we obtain two interesting compatible non trivial 
rational expressions. These rational results for 
real mappings cannot be understood by any obvious Markov's partition,
or symbolic dynamics hyperbolic systems interpretation.

\end{abstract}

\vskip .2cm 

\vskip .2cm 


\vskip .2cm 


\vskip .2cm 


\section{Introduction}

The purpose 
of this paper is to sketch  a classification 
of birational transformations based on various
notions of ``complexity''.
 In previous papers~\cite{topo,zeta,McGuire} an analysis,
 based on the examination of the successive
(bi)rational expressions corresponding to the
iteration of some given
birational mappings, has been performed. When one considers
 the degree $\, d(N)$ of the 
numerators (or denominators) of the corresponding successive
rational expressions for the $N$-th iterate, 
the growth of this degree is (generically) exponential
with $N$ : $\, d(N) \simeq \lambda^N$. $\lambda\, $ has been called
the ``growth complexity''~\cite{complex} and it is closely related to
 the Arnold complexity~\cite{A}.  
A semi-numerical analysis,
 enabling to compute these growth complexities  $\, \lambda$ 
for these birational
transformations, has been introduced in~\cite{topo,zeta}. It has been seen, 
on particular sets of birational transformations~\cite{BoMa95},
that these ``growth complexities'' correspond to
a  remarkable spectrum of {\em algebraic} values~\cite{complex}.

These ``growth complexities'', summing up the (asymptotic) 
evolution of the {\em degree}
of the successive iterates, amount to seeing
these mappings as mappings of (two) {\em complex} variables.
However, when one considers the phase portrait of these mappings,
one also gets some ``hint'' of the  ``complexity''
of these mappings seen as mappings of (two)  {\em real} variables.
In the following, we will consider a 
one-parameter dependent birational  mapping of two variables.
On this very example, it will be seen, considering  phase portraits
corresponding to various values 
of the parameter,  that these ``real complexities''
vary for the different (positive) 
values of the parameter. Two universal (or ``topological'')
measures of the complexities were found to identify~\cite{topo,zeta}, namely the
(asymptotic of the) {\em  Arnold
complexity}\cite{A}  (or growth complexity) 
and the (exponential of the) {\em topological 
entropy}~\cite{topo}. The topological entropy
is associated with the exponential growth $\, h^N$ 
of the number of fixed points (real or complex) of the $\, N$-th iterate 
of the mapping : looking at various phase portraits,
corresponding to different values of the parameter (see below), 
it is tempting to define, in an equivalent way, a ``real topological
entropy'' associated with the exponential growth  $\, h_{real}^N$ 
of the number of {\em real fixed points only} of the $\, N$-th iterate 
of the mapping. This notion of ``real topological
entropy'' would actually correspond to the ``visual complexity'' as
seen on the phase portrait of the mapping.
Such a concept, corresponding
 to the evaluation of the  real complexity  $\, h_{real}\, $ of 
the mapping seen as a mapping bearing on {\em real} variables, would
be less universal: it would have only ``some'' of the remarkable
topological universal properties of 
the topological entropy. Similarly, it is also tempting to 
slightly modify the definition of the {\em Arnold complexity}~\cite{A}. 
The Arnold complexity~\cite{A}, which corresponds (at least for the mappings of
two variables)
to the degree growth complexity~\cite{zeta,McGuire},
 is defined as the number of (real or complex)
intersections of a given (generic and complex) line 
with its $\, N$-th iterate : it is straightforward to similarly
define a notion of ``real Arnold complexity'' 
describing the number of {\em real}
intersections  of a given (generic) {\em real} line 
with its $\, N$-th iterate. This real-analysis concept is, at 
first sight, also very well-suited to
describe the ``real complexity'' of the mapping as it can be seen in the
phase portrait (see Fig.\ref{f:fig2} below). Recalling the identification,  
seen on this one-parameter family
of birational mappings, between the (asymptotic of the)
Arnold complexity and the (exponential of the)
topological entropy~\cite{topo,zeta}, it is natural to 
wonder if this identification
also works for their ``real'' partners, or if, as the common wisdom 
could suggest, real analysis is 
``far less universal'', depending on a lot of 
details and, thus, requires a ``whole bunch'' of 
``complexities'' (Lyapounov dimensions, ...) 
to be described properly. 

In order to see the previous identification
even more clearly, one can also slightly modify the definition
of the dynamical zeta function, introducing
a ``real dynamical zeta function'' associated with the counting
of the real cycles only, and, similarly, 
one can also introduce some ``real Arnold complexity''
generating functions. We will show that several of these 
two ``real'' generating functions  have 
the same singularities. Furthermore we will actually 
conjecture several {\em simple rational expressions} 
for them, yielding, again, 
{\em algebraic values} for the (exponential of the) 
``real topological entropy''. In particular, when
 the  parameter of our family
of birational transformations
 becomes large, we will get an interesting non trivial 
rational expression. These rational results for 
real mappings cannot be simply understood 
by any ``obvious'' Markov's partition,
or symbolic dynamics hyperbolic interpretation.

\section{Growth (Arnold) complexity for a birational mapping}
\label{grArno}
A  {\em one-parameter} family of birational mappings
of  {\em two} (complex) variables
has been introduced in previous papers~\cite{zeta,BoMaRo93c,BoHaMa97}
(see definition (3) in~\cite{zeta}).
This mapping actually originates from a lattice statistical mechanics
framework that will not be detailed
 here~\cite{BoMaRo93c,BoMaRo93a,BoMaRo94}.
 In the following, we will use the extreme simplicity
of this mapping of  {\em two} (complex) variables  to first
 compare two quite universal (topological) 
notions of ``complexity'' namely 
the growth complexity $\, \lambda  $, which measures
the exponential growth of the {\em degree} 
of the successive rational expressions encountered in an iteration
(a notion which coincides with the (asymptotic of the)  {\em Arnold 
complexity}\footnote{
More precisely the Arnold complexity $\, C_A(N)\, $ 
 is proportional (for plane maps) to $\, d(N)$,
 the degree of the $ N$-th
 iteration of the birational mapping which behaves
like $\, d(N) \simeq \lambda^N$. This ``degree notion'' was 
also introduced by A. P. Veselov
in exact correspondence with the general Arnold definition~\cite{A}.
Note that the concept of Arnold complexity {\em is not}
 restricted to two-dimensional maps.}~\cite{A}),
 and the (exponential of the) 
{\em topological entropy}~\cite{AKM65,Ba91}. In section (\ref{Bsome}),
we will go a step further
 and compare, more particularly, 
the notion of {\em ``real'' Arnold complexity} 
versus the notion of   {\em ``real'' topological entropy}.
These two notions will be seen to be suitable to 
describe the properties of the mapping seen as a 
mapping of {\em real} variables.

\subsection{A one-parameter family of birational
transformation}

Let us consider the following 
birational transformation (see (3) in~\cite{zeta})
of two ({\em complex}) variables $\, k_{\epsilon}$,
depending on one parameter $\, \epsilon$ :
\begin{eqnarray}
\label{yz}
 k_{\epsilon} : \qquad \quad
 (y_{n+1}\, , \, \, z_{n+1}) \, \, =  \, \, \, \Bigl(z_n +1 - \epsilon 
\, ,\,  \, \,  \, y_n \cdot \frac{z_n-\epsilon}{z_n + 1} \Bigr)
\end{eqnarray}

In spite of its simplicity, this birational mapping
can, however,  have quite different behaviors
according to
 the actual values of the  parameter $\, \epsilon$.
For example, for $\, \epsilon\, =\, 0$, as well as
  $\epsilon= -1$,  1/2, 1/3 or 1, the
mapping becomes {\em integrable}, whereas
it is not~\cite{BoHaMa97}
 for all other 
values of  $\,  \epsilon$.

Let us now compare, in the following,  two notions of ``complexity''
({\em Arnold complexity versus 
topological entropy})
according to  various values of  $\, \epsilon\, $. 

\subsection{Semi-numerical approach for the growth complexity $\lambda$}
\label{seminum}
The (growth) complexity $\, \lambda $, which measures
 the exponential growth
of the degrees of the successive  rational expressions
one encounters in the iteration of the birational
transformation (\ref{yz}), can  be obtained 
by evaluating  the degrees of the numerators, or equivalently 
 of the denominators,
of the successive (bi)rational expressions obtained in
 the iteration process. One can actually build a {\em semi-numerical
 method}~\cite{topo,zeta} to get
the value of the complexity growth $\lambda$ for any value
of the parameter $\, \epsilon$. The
 idea is to iterate, with the birational transformation
 (\ref{yz}), a generic
{\em rational} initial point $\, (y_0,z_0)\, $ and to
 follow the magnitude
of the successive numerators, or denominators, of the 
iterates.
During the first few steps some accidental simplifications may occur,
but, after this transient regime, the integer denominators (for instance) grow
like $\, \lambda^n$ where $n$ is the number of iterations.
Typically, a best fit of the logarithm of the numerator as a linear function
of $n$, between $\, n=10$ and $\, n=20$, gives the value of $\, \lambda\, $ within
an accuracy of $0.1\%$. Let us remark that an {\em integrable mapping}
yields  a  {\em  polynomial growth} of the calculations~\cite{BoMaRo94} :
the value of the complexity $\lambda$  has to be numerically
very close to $\, 1\, $.
\begin{figure*}
\centerline{
\psfig{file=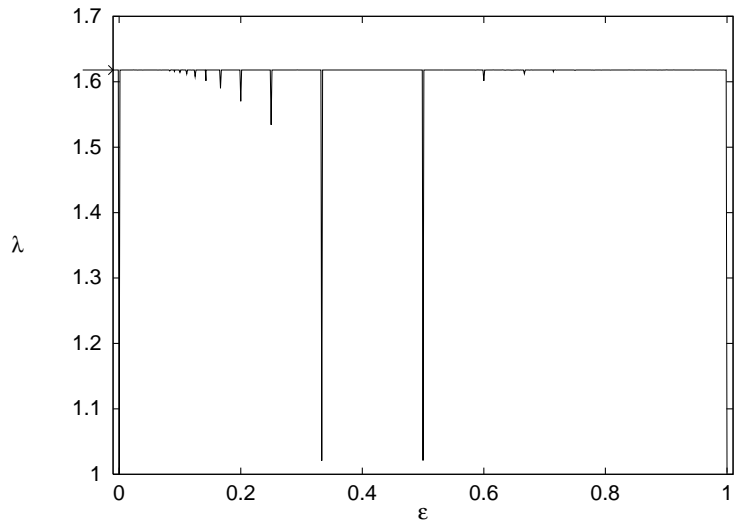}
}
\caption{ Complexity $\lambda$,
 for $k_{\epsilon}$, as a function of $\epsilon$.
\label{f:fig1}
}
\end{figure*}
Fig.\ref{f:fig1} shows
 the values of the complexity growth $\, \lambda$
as a function of the parameter $\, \epsilon$. 
One remarks on Fig.\ref{f:fig1} that all the values  of $\, \epsilon \, $
(except a zero measure set)
give a growth complexity $\, \lambda \simeq 1.618 $.
The calculations have
been performed using an infinite-precision\footnote{
	The multi-precision library gmp (GNU MP) is part of the GNU project.
It is a library for {\em arbitrary} precision arithmetic, operating on
signed integers, rational numbers and floating points numbers.
It is designed to be as fast as possible, both 
for small and huge operands.  The
current version is : 2.0.2. Targeted platforms and Software/Hardware
requirements are any Unix machines, DOS and others, 
with an operating system with reasonable include files and a C compiler.
} C-library~\cite{C}. This semi-numerical analysis~\cite{zeta}
clearly indicates, that, beyond the 
known integrable  values~\cite{BoHaMa97}  of $\, \epsilon $,
 namely $\, -1, \, 0, \,
1/3,\,  1/2,\,  1$, two sets of  values $\{ 1/4,\,  1/5, \, 1/6, \cdots ,\,  1/13 \}$
and $\{ 3/5,\, 2/3\, , \, 5/7 \}$ are
 singled out. This suggests that the growth 
complexity $\, \lambda$ takes lower values than the generic one
on two infinite sequences
of  values of $\epsilon$, namely
$\epsilon = 1/n$  and $\epsilon = (m-1)/(m+3)$ 
for $n$ and $m$ integers such that $n \ge 4$ and
$m \ge 7$ and $\, m$ odd.

\subsection{Generating functions for the degree growth of the
 successive iterates}
\label{Generdegree}
One can revisit all these results using the stability of the
factorization schemes, performing 
exact formal (maple)  calculations.
For instance, one can 
consider, for various values of 
$\, \epsilon  $, the degrees of the successive rational 
expressions one encounters when performing 
the successive iterates, and build various generating
functions\footnote{Similar calculations of generating
functions have been performed~\cite{zeta} using other 
representations of the mapping related
 to $\, 3 \times 3\, $ matrices~\cite{Zittartz}.
These generating
functions, denoted  $\, G_\epsilon( x)$ in~\cite{zeta}, are deduced 
from the existence of  remarkable stable {\em factorization 
schemes}~\cite{zeta,complex,BoMa95}. These
results are in complete agreement with the one given here for the
mapping of two variables (\ref{yz}) (see for instance equations (11), (12), (13)
and (14) in~\cite{zeta}).
}
corresponding to these successive degrees. In particular, having
 singled out a set of values of $\, \epsilon \, $, one
 can revisit these various values,  to see how the generic 
growth complexity  $ \, \lambda \, \simeq \,  \, 1.618   $ gets modified, 
and deduce the degree generating functions, and 
the associated complexity $\, \lambda$, in each case.
Let us denote by $\, G_\epsilon( t)$ 
the  degree generating function, corresponding, 
for some given value of the  parameter $\, \epsilon$,
to the degree of (for instance) the numerator of the $z$ component
of the successive rational expressions obtained in the iteration
process of transformation (\ref{yz}).

At this step, it is worth recalling, again, the notion
of Arnold complexity~\cite{A} which corresponds to 
iterate a given (complex) line and count  the number $\,
A_N\, $ of intersections
of this $N$-th iterate with the initial line.
It is straightforward to see that these
 Arnold complexity numbers
$\, A_N$ are closely linked to these  successive degrees (see for
instance~\cite{topo,zeta}). Actually, if 
one considers the iteration of the
$\, y\, = \, (1-\epsilon)/2\, $ line\footnote{Which is known to be 
a singled-out line for this very mapping~\cite{zeta} (see also, below,
in sections  (\ref{Creal}), (\ref{RAC})).}, the generating function
of the $\, A_N$'s {\em  ``almost'' identifies}
 (often up to a simple $t/(1+t)$ factor)
with the degree generating functions $\, G_{\epsilon}(t)$.
The ``Arnold'' generating functions $\, A_{\epsilon}(t) $,
 and  the degree 
generating functions $\, G_{\epsilon}(t)$, read (up to order 
 fifteen for the ``Arnold'' generating functions and order ten, or eleven, 
for the degree generating functions) :
\begin{eqnarray}
\label{firsteq}
&&A_{\epsilon}(t) \, =\,
\frac t{1+t} \cdot G_{\epsilon}(t)\,  = \frac t{1-t-t^2}\, , 
\qquad \qquad  A_{1/m}(t) \,  = \,
\frac t{1+t} \cdot G_{1/m}(t)\,  = \, \frac t{1-t-t^2+t^{ m + 2}}\, ,
\\ \nonumber
&&A_{(m-1)/(m+3)}(t)\, = \, \,  \,
\frac t{1+t} \cdot G_{(m-1)/(m+3)}(t)\, = \,  
\frac t{1-t-t^2+t^{ m + 2}} \qquad\qquad\qquad
\quad \mbox{for} \quad m\, = 9\, , \, 13\,, \, 17\, , \, 21\, , \,   \ldots \, ,
\\ \nonumber
&&A_{(m-1)/(m+3)}(t)\, = \, 
\frac {t\cdot (1-t^{(m+1)/2})}{1+t} \cdot G_{(m-1)/(m+3)}(t)\, =\,
\frac {t\cdot (1-t^{(m+1)/2})}{1-t-t^2+t^{ m + 2}}
\ \qquad \mbox{for}\quad m\, = 7 \,, \,  \, 11\,, \, 15  \,  , \, \ldots \, ,
\end{eqnarray}
where the expression for $\, G_{\epsilon}(t)\, $ is valid for
$\, \epsilon \, $ generic
and the expressions for
 $\,  G_{1/m}(t)\, $ 
are valid for $\,  m \, \ge \, 4 \, $,
 the  $\, G_{(m-1)/(m+3)}(t)\,$ for  $\,  m \, \ge \, 7  \, $ 
with $\, m $ odd.
One also has for various   integrable value of $ \, \epsilon \, $ :
\begin{eqnarray}
\label{corner}
&& A_{-1}(t) \, = \, \frac t{1+t}\cdot G_{-1}(t) \, = \,
\frac t{1-t^2}\, , \qquad
 A_0(t) \, = \, \frac t{1+t}\cdot G_{0}(t)\,=\, 
\frac t{(1-t)\,(1+t)}\, , 
\nonumber \\
&& A_1(t) \, = \, \frac {t\cdot(1-t)}{1+t}\cdot G_1(t) \, = \,
\frac t{1-t^2}\, ,
\qquad  A_{1/3}(t) \, = \, \frac{t\cdot(1+t)\cdot(1-t)^2}{1+t^4} 
\cdot G_{1/3}(t) \, = \,\frac {t\cdot(1+t)}{1-t^3}\, ,
\\ \nonumber
&& A_{1/2}(t) \, = \, \, \frac t{1+t} \cdot G_{1/2}(t)\, = \,
\frac{ t \cdot (1-t^9)}{(1-t)\cdot(1-t^2)\cdot(1-t^3)\cdot(1-t^5)}\, .
\end{eqnarray}

These various exact generating functions are in agreement with the 
previous semi-numerical calculations. In particular, the first
expression
in (\ref{firsteq}) yields an algebraic value  for $ \, \lambda \, $
 in agreement
with the generic  value of the complexity 
 $ \, \lambda \, \simeq \,  \, 1.618   $ of
  Fig.\ref{f:fig1} (and Fig.1 in~\cite{zeta}).

\section{Dynamical zeta function and  topological entropy}
\label{dynzeta}
It is well known that the periodic orbits (cycles) 
of a mapping $k$ strongly ``encode'' dynamical systems~\cite{Bo72}.
The fixed points of the $N$-th power of the mapping
being the cycles of the mapping itself, their proliferation
with $N$ provides a ``measure'' of chaos~\cite{ASY,Das}. To keep track of
this number of cycles, one can
 introduce the fixed points generating
function
\begin{equation}
\label{genptfixe}
H(t)=\sum_N \# {\rm fix}(k^N) \cdot  t^N 
\end{equation}
where $\, \# {\rm fix}(k^N)) \, $ is the number of fixed points of $\, k^N$,
 {\em real or complex}. This quantity only depends  on the number of
fixed points, and {\em not} on their particular localization. In this
respect, $\, H(t)$ is a {\em topologically invariant
quantity}.
The same information can also be coded in the so-called\footnote{The 
dynamical zeta function has been introduced
by analogy with the Riemann $\zeta$ function, by 
Artin and Mazur~\cite{AM65}.}
{\em dynamical zeta function} $\, \zeta(t)$~\cite{Ba91,Ru78} related
to the generating function $\, H(t)$ by
$\, H(t)  =  t\, \frac{ d}{dt} log(\zeta(t))$.
The dynamical zeta function is defined as follows~\cite{Bo72,AM65,Ru78} :
\begin{equation}
\label{zeta}
\zeta(t)\,  =\, \,  \exp{ \left( \sum_{N=1}^{\infty}{\# {\rm fix}(k^N)}\cdot
 \frac{t^N}{N} \right) }
\end{equation}
The topological entropy $\, \log{h}\, $ is :
\begin{equation}
\label{h}
\log{h}\,  =\, \,  
\lim_{N \rightarrow \infty}{\frac{\log{( \# {\rm fix}(k^N))}}{N}}
\end{equation}
 If the dynamical zeta function is {\em rational},
 $h$ will be the inverse of the pole of smallest modulus
of $H(t)$ or $\zeta(t)$ .
If the dynamical zeta function can be interpreted as the ratio
of two characteristic polynomials of two linear operators\footnote{For
more details on these Perron-Frobenius, or Ruelle-Araki transfer
operators, and other shifts on Markov partition in a symbolic
dynamics
framework, see for instance~\cite{Ru78,Bo73,Bo75,Bo70}.
In this linear operators framework, the {\em rationality} of 
the zeta function, and therefore the algebraicity
of the (exponential of the) topological entropy, 
amounts to having a {\em finite dimensional
representation} of the linear operators $\, A$ and $B$.}
$A$ and $B$,
namely $\zeta(t) = \det(1\, -\, t \cdot B) / \det(1\, -\, t \cdot A) $,
then the number of fixed points  $\, \, \#{\rm fix}(k^N) \, \, $ 
can be expressed from
${\rm Tr}(A^N)-{\rm Tr}(B^N)$. In this case, the poles 
of a rational dynamical zeta function are 
related to the (inverse of the zeroes of the)
characteristic
polynomial of the linear operator $\, A$ only. Since the number of fixed points
 remains unchanged under
{\em topological conjugacy} (see Smale \cite{S67} for this notion),
the dynamical zeta function is also a {\em topologically
invariant function}, invariant under a large set of transformations,
and does not depend on a specific choice of variables. Such
invariances were 
also noticed for the  growth complexity 
$\lambda$. It is thus tempting to make a connection between
the {\em rationality} of the complexity generating 
function previously given, 
 and a possible {\em rationality} of the dynamical
zeta function. We will also compare the singularities
of these two sets of generating functions,namely
the growth complexity $\lambda$
and   $\, h$, the (exponential of the) topological entropy.

\vskip .1cm 
{\bf Some results for the dynamical zeta function  :} Let us now
 get the expansion of the dynamical zeta
 function of the mapping
$\, k_\epsilon$, for generic
 values of $\epsilon$. We can first concentrate on the specific\footnote{
Another generic value of $\, \epsilon$, close to the  1/2 value 
 where the mapping is
integrable~\cite{BoHaMa97}, namely  $\, \epsilon\, =\, 13/25\, =\,
0.52$,
has been analyzed in some detail in~\cite{zeta}.
For this value  $\, \epsilon\, =\, \, 0.52$,  the
enumeration of the number of fixed points, $n$-cycles
and the actual status of these fixed points
(elliptic, hyperbolic, points ...)
are given in~\cite{zeta}.}, but arbitrary,
value $\epsilon \, =\, 21/25\,$. Of course,  there is nothing particular
with this specific $\, \epsilon\, =\, 21/25 $
value : the same calculations have been performed 
for many other {\em generic} values of  $\, \epsilon\,$
yielding the same number of (complex) fixed points
and, thus, the same dynamical zeta function. The total 
number of fixed points of $\,k_{\epsilon}^N$,
for $\,N\, $ running from $\, 1$ to $\, 14$, yields, up to order fourteen,
the following expansion for 
the generating function $\, H(t)$ of the  number of fixed points  :
\begin{eqnarray}
\label{G}
&&H_{\epsilon}(t) \, = \,\, H_{21/25}(t) \, = \,\,\,
 t+t^2+4\,t^3+5\,t^4\,+11\,{t}^{5}+16\,{t}^{6} 
 +29\,t^7 +45\,{t}^{8}+76\,{t}^{9}
 +121\,{t}^{10}+199\,{t}^{11} \nonumber \\
&& \qquad \qquad \qquad  \qquad \qquad  \,+320\,t^{12}\, +521\,{t}^{13}\,
 +841\,{t}^{14}\, + \ldots 
\end{eqnarray}
This expansion coincides with the one of the {\em rational}
function\footnote{Valid for generic values of $\, \epsilon $, up to 
some algebraic values of $\, \epsilon $ corresponding to cycle-fusion
mechanism see (\ref{previous}) and  (\ref{moreprevious}) below and see~\cite{McGuire}.} :
\begin{eqnarray}
\label{GG}
H_\epsilon(t) \, = \,  \, 
{\frac {t \cdot \left (1 \, + \, t^2 \right )}
{\left (1-t^2\right ) \cdot \left (1-t-t^2\right )}}
\end{eqnarray}
which corresponds to a very  {\em simple\footnote{As far as {\em symbolic dynamics}
is concerned, one can associate,
to a dynamical zeta function like (\ref{conjec}),
a clipped Bernoulli shift with the ``pruning rule'' to
forbid substring $-11-$ (that is 1 must be always followed by 0)
in any sequence of 0 and 1. However, constructing the Markovian
 partitions
(if any), yielding this simple pruning rule for the symbolic dynamics,
remains to be done.}
 rational} expression for
the dynamical zeta function :
\begin{eqnarray}
\label{conjec}
\zeta_\epsilon(t)  \, \,  = \, \,  \,  {{1\, -t^2 } \over {1\, -t\, -t^2}} 
\end{eqnarray}
 An alternative
 way of writing the dynamical zeta functions
 relies on the decomposition of the 
fixed points into {\em irreducible  cycles} :
\begin{eqnarray}
\label{Weil}
\zeta_\epsilon(t) \,\,\, = \,\,\,\,\, {{1} \over {(1-t)^{N_1}}} \cdot 
{{1} \over {(1-t^2)^{N_2}}} \cdot
 {{1} \over {(1-t^3)^{N_3}}} \,\,  \cdots\,\, 
 {{1} \over {(1-t^r)^{N_r}}}\, \, \cdots\,\,  
\end{eqnarray}
For generic  values of $\epsilon$,
one gets the following numbers of irreducible cycles :
 $\, N_1\, = \, 1\, , \, N_2\, = \, 0\,, \,  N_3\, = 1\, , 
\,  N_4\, = 1\, ,\,  N_5\, = 2\, ,  \,  N_6\, = 2\, , 
\,  N_7\, = 4\, , \,  N_8\, = 5\, ,$$ \,  N_9\, = 8\, , 
\,  N_{10}\, = 11\, , \,  N_{11}\, = 18\, , \, \cdots $
It has been  conjectured in~\cite{zeta} that 
{\em the  simple rational expression}
~(\ref{conjec}) {\em is the actual  expression 
of the dynamical zeta function for
 any generic value of $\epsilon$} (up to some algebraic
values of $\, \epsilon$, see below and in section (\ref{Creal})). 
Similar calculations have been performed for the other values of 
 $\epsilon \, $ that have been singled out 
in the semi-numerical analysis~\cite{McGuire}.  For  the 
non generic values of $\epsilon$,  $ \epsilon\, = \, 1/m\, $
with  $\,m \, \ge \, 4 $, 
we have obtained expansions compatible 
with the following {\em rational} 
expression :
\begin{eqnarray}
\label{zetamtexte}
\zeta_{1/m}(t)  \quad = \quad
 {{1\, -t^2 } \over { 1\, -t\, -t^2\, + t^{m+2}}} 
\end{eqnarray}
For the other non generic values, namely 
 $ \epsilon\, = \, (m-1)/(m+3)\, $ with $\, m\, \ge 7 $ odd,
the expansions are not large enough to conjecture 
a single formula valid for any $\, m$. For $m=7$ (namely 
$\, \epsilon \, = \, 3/5$) one actually gets a dynamical zeta function
 given by (\ref{zetamtexte}) for $m\, = \, 7$ and this might also 
be the case for $m\, = \, 11\, , \, 15 \, , \, \cdots$. For $\, m \, =
 \, 9\, , \, 13\, , \, \cdots\, $ the expansions are in agreement with
a  $\, 1\, -t\, -t^2\, + t^{m+2}\, $ singularity.
 Comparing the various rational expressions in~(\ref{firsteq})
 corresponding to generic, and non-generic, values of $\, \epsilon$,
with  (\ref{conjec}), and (\ref{zetamtexte}), respectively, one sees 
that the singularities (poles) of the dynamical zeta function
happen to  {\em coincide} with the poles of the 
generating functions of the growth complexity $\, \lambda$, for all
 the values of $\epsilon$.
In particular, the  growth complexity  $\, \lambda \, $ 
and $\, h$, the exponential of the topological entropy, are always {\em equal}.

Let us just mention, here, that the
modification 
of the number of fixed points, from the 
``generic'' values of $\, \epsilon \, $ to the particular values
($1/m$, $(m-1)/(m+3)$),
corresponds to  the {\em disappearance
of some cycles} which become singular points (indetermination of the
form $0/0$). These mechanisms will be detailed in~\cite{Lyapou}.
Actually the ``non-generic''
values of $\epsilon\, $, like $\epsilon \, = \, 1/m$, 
correspond to such a  ``disappearance
of  cycles'' mechanism which modifies the denominator of the rational
generating functions, and, thus, the topological entropy and the growth
complexity $\lambda$. In contrast, there actually exists for $k_{\epsilon}$,
{\em other singled-out values} of $\epsilon\, $, like $\epsilon \, = \,3$
for instance, which correspond
 to {\em fusion of cycles} (see section (\ref{Creal})) :
in the $\epsilon \, \rightarrow  \,3\, $ limit, the order three cycle tends
to coincide with the order one cycle, which amounts
 to multiplying the dynamical zeta function (\ref{conjec})
by $1-t^3$. Such
 ``fusion-cycle'' mechanism {\em does not modify} the
denominator of the rational functions, and thus, 
the topological entropy, or the growth complexity $\, \lambda$, {\em  remain
unchanged}.

{\bf To sum up :} Considering a (very simple)
 one parameter-dependent birational mapping
 of {\em only two
(complex) variables}, we have deduced
 an exact identification between the (asymptotic of the)
Arnold complexity, that is the growth complexity $\, \lambda $,
and the (exponential of the) topological entropy for {\em all} the various
$\, \epsilon$ cases (generic or not). This identification can be
understood heuristically~\cite{zeta}.
As a byproduct, one finds that these two
complexities correspond, in this very example, to 
{\em simple algebraic numbers}. 

\vskip .3cm 
{\bf A canonical degree generating function :} This identification result
 is not completely surprising :
the dynamical zeta function is a quite 
``universal'' function, invariant under a large set of 
 {\em topological conjugaisons}~\cite{S67}, and  the 
concept of Arnold complexity (or the degree growth complexity
$\, \lambda$)
also has the same ``large'' set of
(topological and projective) 
invariances~\cite{A}. 

Actually, as far as degree generating
functions are concerned, it is natural to introduce, instead of 
some  generating
functions of the degrees of the numerator of the $z$ component of the
$N$-th iterate, a more ``canonical'' degree generating
function $\, G^{Hom}_{\epsilon}(t)\,$ associated with
the birational mapping (\ref{yz}) written in a {\em homogeneous
way} (see the bi-polynomial mapping (4) in~\cite{zeta}).
Iterating  (\ref{yz}), written in a homogeneous
way, and, factoring out
at each iteration step the gcd's, one gets a new degree
generating function $\, G^{Hom}_{\epsilon}(t)\,$ 
well-suited, at first sight, to 
describe such large (topological and projective) invariances.
A simple calculation shows that this
 (projectively well-suited) degree
generating function reads (for generic $\, \epsilon$) :
\begin{eqnarray}
\label{ghom}
G^{Hom}_{\epsilon}(t)\, \,= \, \, \, \,
 {{ 1} \over { (1-t) \cdot (1-t-t^2)}} \,\,\, \, 
\end{eqnarray}
For the $\, \epsilon \, = \, 1/m\, $ particular values, and for the
two integrable values, $\epsilon \, = \, 1/2$
and  $\epsilon \, = \, 1/3$, one gets respectively :
\begin{eqnarray}
\label{1surmHom}
&&G^{Hom}_{1/m}(t)\, \,= \, \, \, \,
  {{ 1-t^{m+3}} \over { (1-t) \cdot (1-t-t^2+t^{m+2})}}  \\
&&G^{Hom}_{1/2}(t)\, \,= \, \, \, \, {{1-t^9 } \over {
  (1-t)^2 \cdot (1-t^3) \cdot (1-t^5)}} \, + \, {{t^2 \cdot (1-t^6) } \over 
{(1-t)^2 \cdot (1-t^2) \cdot (1-t^5)}}  \\
&&G^{Hom}_{1/3}(t)\, \,= \, \, \, \,
 {\frac {1-{t}^{6}}{\left (1-{t}^{3}\right )\left (1-{t}^{2}\right )
\left (1-t\right )^{2}}}\, +\, 
{\frac {{t}^{4}}{\left (1-{t}^{3}\right )\left (1-t\right )^{2}}}
\end{eqnarray}
Since the expansions for the infinite set of values of the form 
$\, (m-1)/(m+3)\, $ for $\, m \, \ge \, 7$, can only be performed up
to order eleven (or twelve), it is difficult to ``guess'' any  
expression valid for any $\, m \, $ like (\ref{1surmHom}). 
Recalling the results (\ref{firsteq}) given
 in section (\ref{Generdegree}) for the degree
growth generating functions, one may suspect that, among these 
$\, (m-1)/(m+3)\, $ for $\, m \, \ge \, 7$ values,
one should make a distinction between 
 $\, m \, = \, 7\, , \,  11\,, \, \, 15\, , \, \ldots\, $ on one side, and 
 $\, m \, = \, 9\,, \, 13\,,\, 17 \, , \, \ldots\, $ on the other
side. In fact, up to order eleven, all our calculations for
various $\, (m-1)/(m+3)\, $ 
values for $\, m \, \ge \, 7$ ($\, \epsilon \, = \, 3/5\, , \, 2/3\, ,
 \, 5/7\, , \, 3/4\, , \, 7/9\, , \, 4/5\, , \, \ldots\, $) 
are in agreement 
with a general equality between $\, G^{Hom}_{(m-1)/(m+3)}(t)\,$
 and $\, G^{Hom}_{1/m}(t)$. More details are available in Appendix~A.

One gets simpler expressions for the integrable values $\epsilon \,
= \, 0\, , \, 1$ and $\epsilon \,
= \, -1$:
\begin{eqnarray}
G^{Hom}_{0}(t)\, \,= \, \, \, \,G^{Hom}_{1}(t)\, \,
= \, \, \, \, {{1+t^2} \over {(1-t)^2}}
\, , \qquad \qquad \hbox{and :} \qquad \qquad 
G^{Hom}_{-1}(t)\, \, \,  = \, \, \, \, \, {{1} \over {1-t}}
\end{eqnarray}
One remarks that $\, G^{Hom}_{\epsilon}(t)\,$ verifies
the simple functional equation 
 $\, G^{Hom}_{\epsilon}(t)\, + \, G^{Hom}_{\epsilon}(1/t)\, = \, 1$,
for $\, \epsilon \, = \, -1\, , \, 1/2\, , \, 1/3$,  and 
 $\, G^{Hom}_{\epsilon}(t)\, = \, G^{Hom}_{\epsilon}(1/t)\, $
for $\, \epsilon \, = \,0\, , \,  +1$.

For $\, \epsilon \, = \, 1/2\, $
 we have not written any dynamical zeta
function $\, \zeta_{1/2}(t)\, $ because, for such an integrable 
birational mapping,
there exist, at (almost) any order  $\,N $ of iteration, an
 {\em infinite number} of fixed
 points of order  $\,N $ (all the points 
of some elliptic curves~\cite{BoHaMa97}) and, therefore, 
our previous ``simple'' definition (\ref{zeta})
for the dynamical zeta
function is not valid anymore. 
\vskip .3cm
\vskip .1cm
{\bf A possible universal relation.} One can imagine many simple relations
between the ``canonical''
degree generating function, $\,
G^{Hom}_{\epsilon}(t)$, and the dynamical zeta function,
 $\, \zeta_{\epsilon}(t)$. For instance,
 for generic $\, \epsilon $, one gets (among many ...) the relation
$\, (1-t)\cdot (1-t^2) \cdot \, G^{Hom}_{\epsilon}(t)\, \, \,= \, \,
 \zeta_{\epsilon}(t)$, however this relation 
is not anymore valid for $\epsilon \, = \, 1/m$. 
One would like to find 
a ``true universal'' relation between  $\, \zeta_{\epsilon}(t) \,$
and  $\, G^{Hom}_{\epsilon}(t)$, that is a relation
{\em independent} of $\, \epsilon\, $ (generic or non-generic).
In order to achieve this goal one may imagine to barter
 $\, \zeta_{\epsilon}(t) $,
and  $\, G^{Hom}_{\epsilon}(t)$, for {\em projectively well-suited 
generating functions} taking into
account the point at $\, \infty$, namely  
 a dynamical 
zeta function taking into
account the fixed point at $\infty$ 
(see (52) in~\cite{McGuire}), $\, \zeta^{(\infty)}(t)  $,
and  $\, G^{Hom}_{\infty}(\epsilon, t)\, $ 
defined as follows :
\begin{eqnarray}
 \zeta_{\epsilon}^{(\infty)}(t) \, = 
\, \,  {{ \zeta_{\epsilon}(t)} \over {1-t}}
\qquad \qquad  \mbox{and :} \qquad \qquad 
 G^{Hom}_{\infty}(\epsilon, t)\,\,\, \, \,
= \, \, \, \,G^{Hom}_{\epsilon}(t)\,+\, \,  {{t} \over {1-t}}
\end{eqnarray}
One verifies immediately that the relation :
\begin{eqnarray}
\label{universal}
  G^{Hom}_{\infty}(\epsilon, t)\,\,\, = \, \,  (1+t) \cdot
\zeta_{\epsilon}^{(\infty)}(t) \, \qquad \hbox{or equivalently :}
\qquad 
 (1+t) \cdot \zeta_{\epsilon}(t)\,\, = \,  \, \, (1-t)  \cdot
G^{Hom}_{\epsilon}(t)\, + \, \, t
\end{eqnarray}
is {\em actually verified for generic} values of $\, \epsilon$,
as well as, for the {\em non-generic 
values}  of the form $\, \epsilon\, = \, \,  1/m$,
 and also some {\em non-generic 
values}  of the form $\, \epsilon\, = \, \,  (m-1)/(m+3)$ (see Appendix~A).
A similar relation for the two-parameters
family of birational transformations
depicted in~\cite{topo,zeta,McGuire}
will be detailed elsewhere.
Relation (\ref{universal}) should give some hint 
for a true mathematical proof of the 
relation between Arnold complexity and topological entropy.
\vskip .3cm
{\bf Remark :} Recalling the 
``Arnold'' generating functions $\, A_{\epsilon}(t) $, (see
(\ref{firsteq})), which identifies ``most of the time'' (namely
$\, \epsilon\, $ generic, $\, \epsilon\,=1/m $,
$\, \epsilon\,=(m+1)/(m+3) $ for $m\, =\, 9\, , \, 13\, , \, 17\, , \,
\ldots$)
with  the new well-suited generating
function  $\, G^{Hom}_{\infty}(\epsilon, t)$,  up to 
a simple multiplicative factor $\, t/(1+t)^2$, one can rewrite, 
for $\, \epsilon\, $ generic and $\, \epsilon\,=1/m $, 
relation (\ref{universal}) as :
\begin{eqnarray}
\label{universal2}
t \cdot \zeta_{\epsilon}(t) \, = \, \, \, (1-t^2) \cdot
A_{\epsilon}(t) 
\end{eqnarray}
\section{Real dynamical zeta function and  real topological entropy}
\label{RDZ}
As far as the growth complexity $\, \lambda\, $ is concerned, 
the generic values of  $\, \epsilon \, $ (that is the values 
different from the previous $1/m$, $(m-1)/(m+3)$
singled-out values) are all on the same ``complexity footing'' 
(see the previous Fig.\ref{f:fig1}). This is clearly confirmed
by the exponential {\em growth of the computing time } 
during the iteration process,
which seems to be similar for all these 
 values (and clearly {\em smaller}
 for the  $\, 1/m$, $(m-1)/(m+3)$
particular values). It is however worth noticing  that these 
generic values, which are all on the same  $\, \lambda $-footing, 
clearly yield phase portraits which are 
quite different and, obviously, correspond
to {\em drastically different} ``visual complexities'' of the 
 phase portrait of the mapping. 
This ``visual complexity'' corresponds to the 
(exponential) growth of the
number of (real) fixed points of the mapping seen as
a mapping {\em bearing on two real variables}.
The previous definitions of the dynamical zeta functions $\,
\zeta_{\epsilon}(t)\, $ and of the
generating function $\, H_{\epsilon}(t) $ counting the number of fixed
points, can be straightforwardly modified to describe
the counting of {\em real fixed points} :
\begin{eqnarray}
\label{defreal}
H_{real}(t)\,= \, \,\sum_N  H^{R}_N \cdot t^N  \,= \, \,
 t\cdot {{d} \over {dt}} log(\zeta^{real}(t))\, ,
\quad \quad \quad  \hbox{where :} \quad \quad \quad 
\zeta^{real}(t) \, \, = \, \, \, \sum_N  z^{R}_N \cdot t^N
\end{eqnarray}
where the number of {\em real} fixed points  $\, H^{R}_N\, $ 
grow exponentially with the number $\, N\, $ of iterates,
like $\,  \simeq \, \, h_{real}^N$. A quick examination of various 
phase portrait for various 
``generic values'' of the parameter $\, \epsilon\, $
seems to indicate quite clearly that this ``real topological entropy''
$\, \log( h_{real}) \, $ {\em varies with}  $\, \epsilon $,
in contrast with the ``usual'' topological entropy $\, \log(h)$.
An obvious inequality is :  $\, \,h_{real}  \le \, h$.
\subsection{``Phase portrait gallery''}
\label{gallery}
Let us give here various phase portraits corresponding to different
(generic except the first one) values of $\, \epsilon$.
Note the different 
scale for the frame of these various phase portraits.
 For most of
the  phase portraits below around
300 orbits of length 1000, starting from randomly
chosen points inside the frame\footnote{With a special 
non-random treatment of the regular elliptic 
parts of the phase portraits.}, have been generated
(only points inside the frame are shown) :
\begin{figure*}
\centerline{
\psfig{file=Fig2.eps,height=6.5cm,width=8cm}
\psfig{file=Fig3.eps,height=6.5cm,width=8cm}
}
\caption{Phase portrait of $k_\epsilon$ for 
$\epsilon=1/100$ (left) and for $\epsilon=9/50$ (right). 
\label{f:fig2}
}
\end{figure*}
\twocolumn
\begin{figure*}
\psfig{file=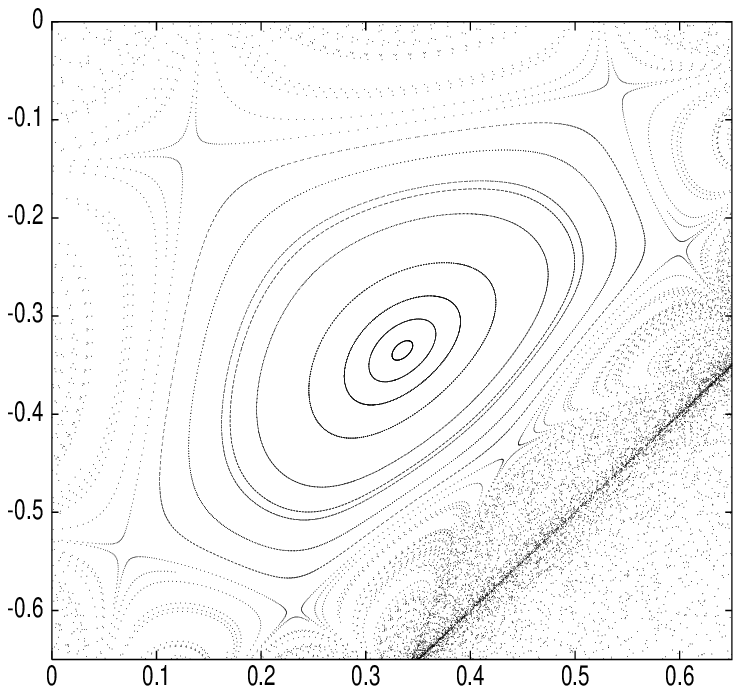,height=6.5cm,width=8cm}
\caption{Phase portrait of $k_\epsilon$ for 
$\epsilon=33/100$.
\label{f:fig4}
}
\end{figure*}
\begin{figure*}
\psfig{file=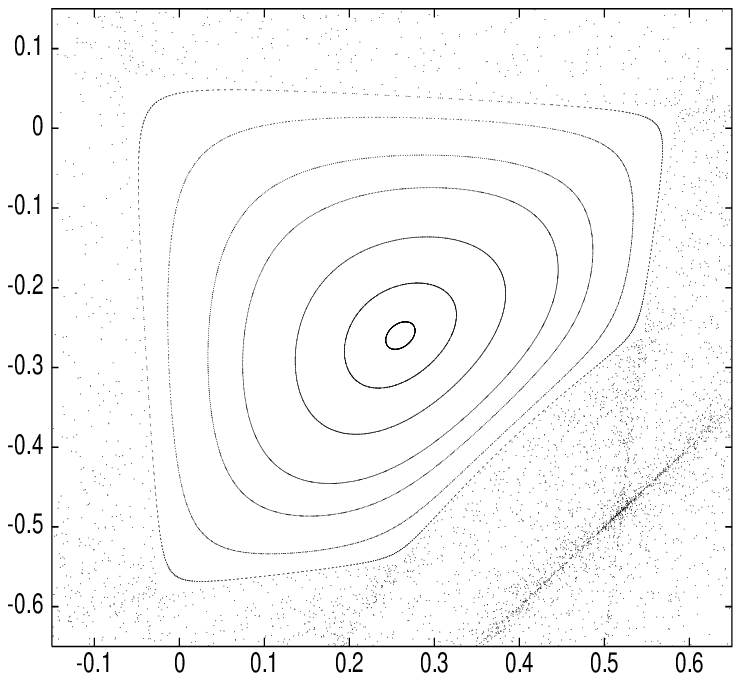,height=6.5cm,width=8cm}
\caption{Phase portrait of $k_\epsilon$ for 
$\epsilon=48/100$. 
\label{f:fig6}
}
\end{figure*}
\begin{figure*}
\psfig{file=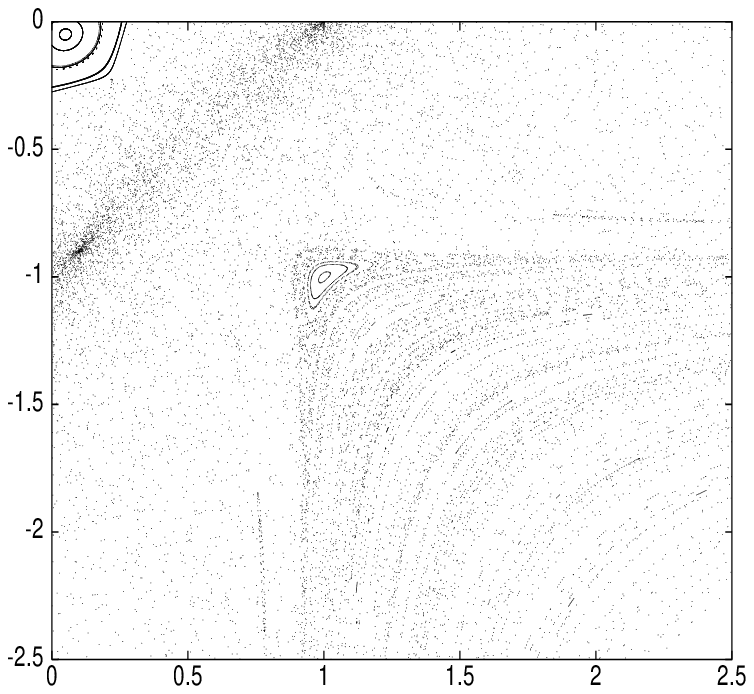,height=6.5cm,width=8cm}
\caption{Phase portrait of $k_\epsilon$ for 
$\epsilon=9/10$. 
\label{f:fig8}
}
\end{figure*}
\begin{figure*}
\psfig{file=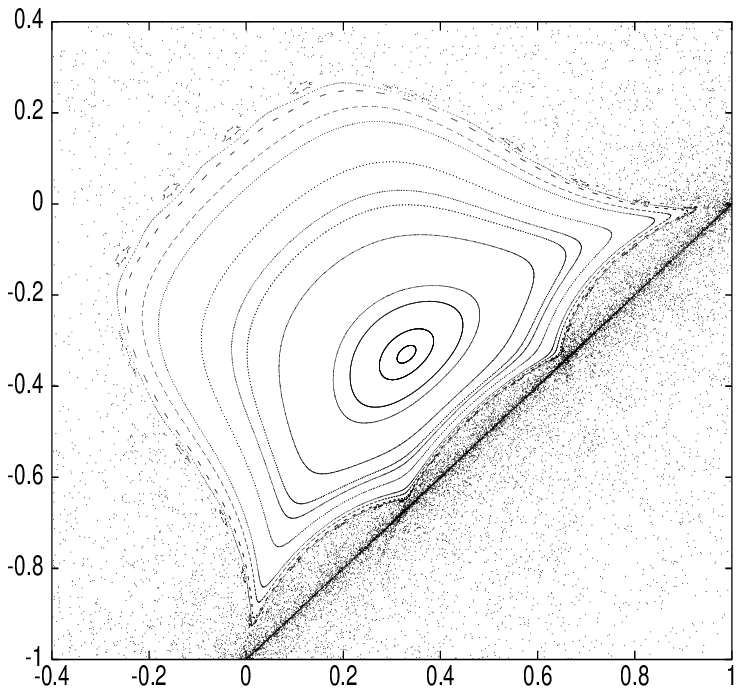,height=6.5cm,width=8cm}
\caption{Phase portrait of $k_\epsilon$ for 
$\epsilon=34/100$. 
\label{f:fig5}
}
\end{figure*}
\begin{figure*}
\psfig{file=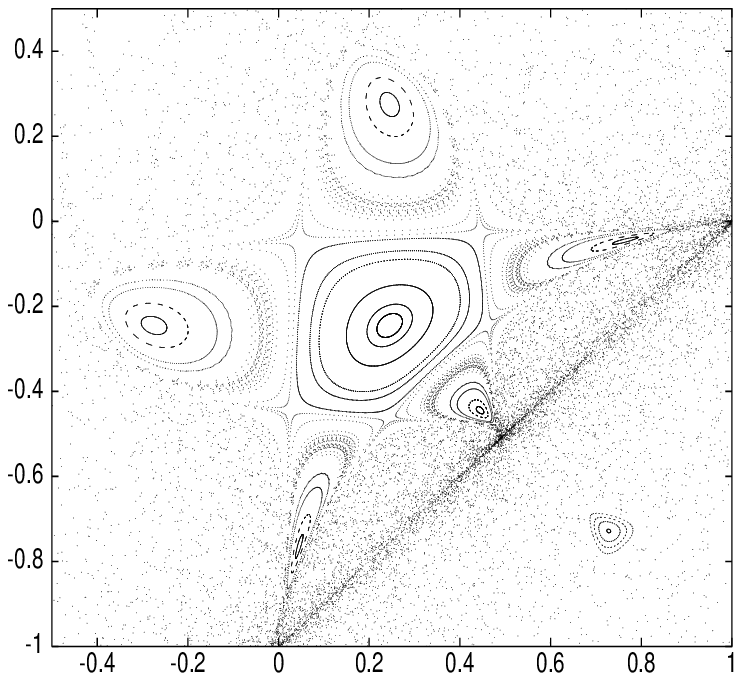,height=6.5cm,width=8cm}
\caption{Phase portrait of $k_\epsilon$ for 
$\epsilon=51/100$. 
\label{f:fig7}
}
\end{figure*}
\begin{figure*}
\psfig{file=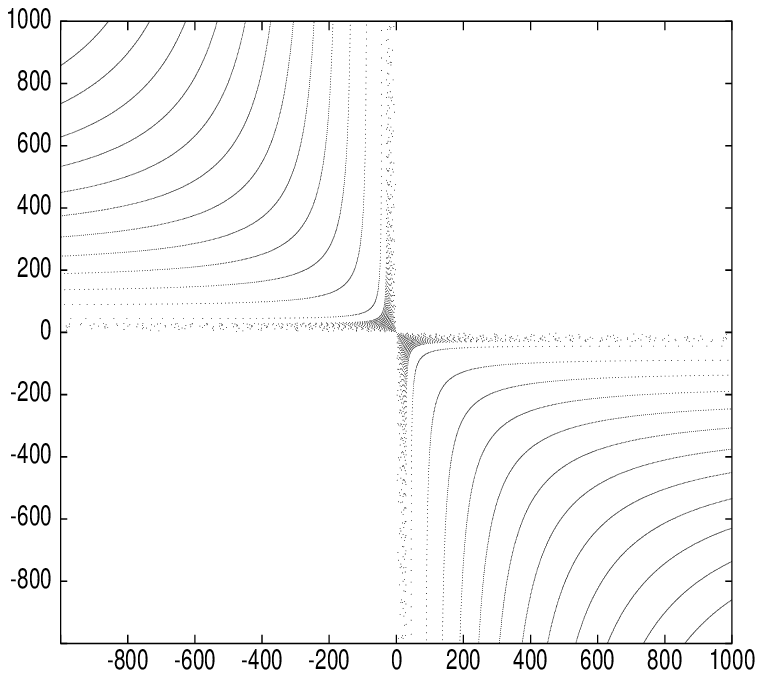,height=6.5cm,width=8cm}
\caption{Phase portrait of $k_\epsilon$ for 
$\epsilon=11/10$. 
\label{f:fig9}
}
\end{figure*}
\onecolumn
\begin{figure*}
\centerline{
\psfig{file=Fig8.eps,height=6.5cm,width=8cm}
\psfig{file=Fig9.eps,height=6.5cm,width=8cm}
}
\caption{Phase portrait of $k_\epsilon$ for 
$\epsilon=3$ (left) and for $\epsilon=15$ (right).
\label{f:fig10}
}
\end{figure*}

On these various phase portraits one sees
that  one 
gets, near the integrable value $\, \epsilon \, \simeq  \, 1/3$,
 quite different phase portraits  which seem,
however, to have roughly the same number of (real) fixed points 
(see Fig.\ref{f:fig4} and Fig.\ref{f:fig5}).
On these various phase portraits one also 
sees, quite clearly, that the
number of fixed points seems to 
{\em decrease} when $\, \epsilon\, $ 
crosses the integrable value $\, \epsilon \, = \, 1/2$
 and $\, \epsilon \, = \, 1$,
going, for instance, from  $\, \epsilon \, = \, .48$ to $ \epsilon \, =
\, .51$ (see Fig.\ref{f:fig6} and Fig.\ref{f:fig7})
or, from $\, \epsilon \, = \, .9  \,$ to $ \, \epsilon \, =
\, 1.1$ (see Fig.\ref{f:fig8} and Fig.\ref{f:fig9}). These results 
will be revisited in a
forthcoming section (see section (\ref{RAC})).
Of course, exactly on the integrable value $\, \epsilon \, = \, 1$,
the phase diagram corresponds to a (simple)
 foliation of the two-dimensional
parameter space in {\em (rational) curves} (linear
 pencil of rational curves,
see~\cite{BoHaMa97}) :
\begin{eqnarray}
\label{foliaepsi1}
\Delta(y,z,1)\, \,=\,\Bigl({\frac {y \, z}{y-z-1}}\Bigr)^2 \, 
= \, \, \rho\, , \qquad \qquad \hbox{or
equivalently :} \qquad \qquad 
{{ y \, z} \over {y-z-1}} \, = \, \, \pm \, \rho^{1/2}
\end{eqnarray}
where $\, \rho$ denotes some constant.
For the other integrable values one also has either a
linear pencil of rational
curves, namely $\, y \, z \, = \, \rho\, \, $ for $\, \epsilon \, = \, 0$,
as well as  :
\begin{eqnarray}
\label{foliaepsmoins1}
\Delta(y,z,-1)\, \,=\, \,{\frac{1}{(1+z-y)^2}} \, = \, \, \rho 
 \qquad \qquad \hbox{or
equivalently :} \qquad \qquad 
(y-z)\cdot (y-z-2)\, \, = \, \,\,  {{1} \over {\rho}}\, -\, 1
\end{eqnarray}
 for $\, \epsilon \, = \, -1$, or a linear pencil of
 {\em elliptic curves} for $\epsilon \, = \, 1/2$, namely :
\begin{eqnarray}
\label{foliaepsi1sur2}
\Delta(y,z,1/2)\, \,=
\,\, \,{\frac{(1\,+z\, +2 \, y\, z)\cdot (1\, -y\, +\, 2\, y\,
z)\cdot (1\, +z\, -y\, -\, 2 \, y\, z )}{(1+z-y)^2}} \,\, \,  = \, \,\, \,  \rho
\end{eqnarray}
and also :
\begin{eqnarray}
\label{foliaepsi1sur3}
\Delta(y,z,1/3)\, \,=\, \, \, \, {\frac{(5+3 z-3 y\, +9 \, y\,
z)\cdot (1\, -z\, -y\, +3 \,  y\, z)\cdot 
(1+z-y\, -3 \,  y\, z) \cdot 
(1\, +z\, +y\, +3 \, y\, z)}{(1\, +z\, -y)^2}} \nonumber
 \, \,   \,= \, \,  \, \rho \,
\end{eqnarray}
for $\epsilon \, = \, 1/3$. One also remarks that 
 $\, \epsilon \, = \, 3$, which corresponds to
the generic $\, \lambda \, \simeq \, 1.618  \, $
growth complexity, also yields a {\em remarkably regular} phase portrait,
``visually'' similar to a foliation 
of the two-dimensional
parameter space in curves, suggesting a 
``real topological complexity''  $\, h_{real}\, $ very close, or even equal to
$\, 1$. This fact will also be revisited in 
the next section (see section (\ref{RAC})).
In order to describe, less qualitatively, 
the  ``real topological complexity'' $\, h_{real}\, $
as a function of the parameter $\epsilon$, we have calculated 
in section (\ref{Bsome}), the 
first (ten,  eleven or even twelve)  coefficients  
of the expansions of $\, H_{\epsilon}^{real}(t)$, and 
of the ``real dynamical function'' $\, \zeta_{\epsilon}^{real}(t)$, 
for various values of $\, \epsilon$.

\subsection{Number of real fixed points as a function of $\, \epsilon$.}
\label{Creal}
Let us now try to understand why (and how)  
$\, h_{real}\, $ varies as a function
of   $\,\epsilon$, and why some other values of 
 $\, \epsilon$, like $\, \epsilon \, = \, 3$, 
 different from the previous $\, 1/m\, $ and $\, (m-1)/(m+3)$
non-generic values, seem to play a special role. The method to get 
the fixed points of the $N$-th iterate of $\, k_{\epsilon}$ 
has been detailed in  previous papers~\cite{topo,zeta}. Let
us just mention here that, due to the symmetries of
 this mapping,  there exist two singled out lines, namely 
$\, y\, = \, (1-\epsilon)/2\, $ and $\, y\, = \,-z $,
playing a key role in classifying all these fixed points~\cite{Alg}.
The fixed points  of $\, k_{\epsilon}^N$
belong to $\, N$-cycles (if $\, {\cal M}\, $ 
is  a fixed point  of $\, k_{\epsilon}^N$
then 
 $\, k^p_{\epsilon}(\cal{M}) \,$ is 
also a fixed point  of $\, k_{\epsilon}^N$, 
for $p \, = \, 1\, , \,  \ldots \, , \, N-1$).
It can be seen   that 
{\em there always exist} a fixed point in these $N$-cycles
which belongs, either to line $\, y\, = \, (1-\epsilon)/2 $, 
or to the  $\, y\, = \,-z $ line. We call 
the fixed points, corresponding to line 
 $\, y\, = \,-z $, the ``P-type'' points and the ones, 
corresponding to line 
$\, y\, = \, (1-\epsilon)/2 $,
the ``Q-type'' points~\cite{Lyapou}.
For $\, N \ge 9$,  
other $N$-cycles with no points lying on these two lines
do occur: other 
remarkable sets  occur like
 $\, y\, + \bar{z} = \,0 $ (see~\cite{zeta,McGuire}). 
 We call 
the fixed points, corresponding this ``remaining'' set of
points, the points of the ``R-type''~\cite{Lyapou}.

One can use these localization properties to
get, very quickly, a subset of all the fixed points, namely, for
instance, the ``Q-type'' fixed points 
(one representant in the  $\, N$-cycle 
belongs to line $\, y\, = \,
(1-\epsilon)/2 $). These calculations can be performed quite
efficiently since one can eliminate the $y$ variable
($\, y\, = \, (1-\epsilon)/2\, $) and, thus,
reduce the calculations to
looking for the roots (real or not) of an
$\epsilon$-dependent
 polynomial
in this remaining $z$ variable. One  gets, for the first values
of $N$, the following polynomial
expressions relating $z$ and $\epsilon$ :
\begin{eqnarray}
\label{Qtype}
&&Q_1(z,\epsilon) \, = \, \, 2 \cdot z - (\epsilon -1)
 \, = \, 0 , \qquad \qquad 
Q_3(z,\epsilon) \, = \,\, z\, -(\epsilon -2)\,\, = \, 0 , \qquad \nonumber \\
&& Q_5(z,\epsilon) \, = \, \, (3 \,\epsilon -1) \cdot z^2 \, - \, 2 \cdot
(\epsilon -3)
\cdot (2 \,\epsilon -1) \cdot z\, +\, (\epsilon^3\, -5 \, \epsilon^2
+10 \, \epsilon\, -4)\, = \, 0 \, ,\, \, \,  \ldots 
\end{eqnarray}
It is  easy to see that
the number of real roots  $z$, of one of these $\, Q_N(z,\epsilon)\, = \,
0$ conditions, varies 
with $\, \epsilon $ by intervals.
The changes of this number of 
real roots take place at
{\em algebraic} values of $\, \epsilon $ (resultant 
of  $\, Q_N(z,\epsilon)\,$ in $z$). The details of the calculations, and a
  description of these polynomials, will be given
elsewhere~\cite{Lyapou}.
The number of the fixed points of the ``$Q$-type'' (see~\cite{Alg}) is,  
thus, a function of $\epsilon$ constant by interval, the limits of the
intervals corresponding to some algebraic values (resultants 
deduced from the $ \,Q_N$'s by eliminating $z$).
For illustration, let us just plot here the real 
roots $\, z$, as a function of $\, \epsilon$,
for $\, Q_{10}$ :
\begin{figure}
\psfig{file=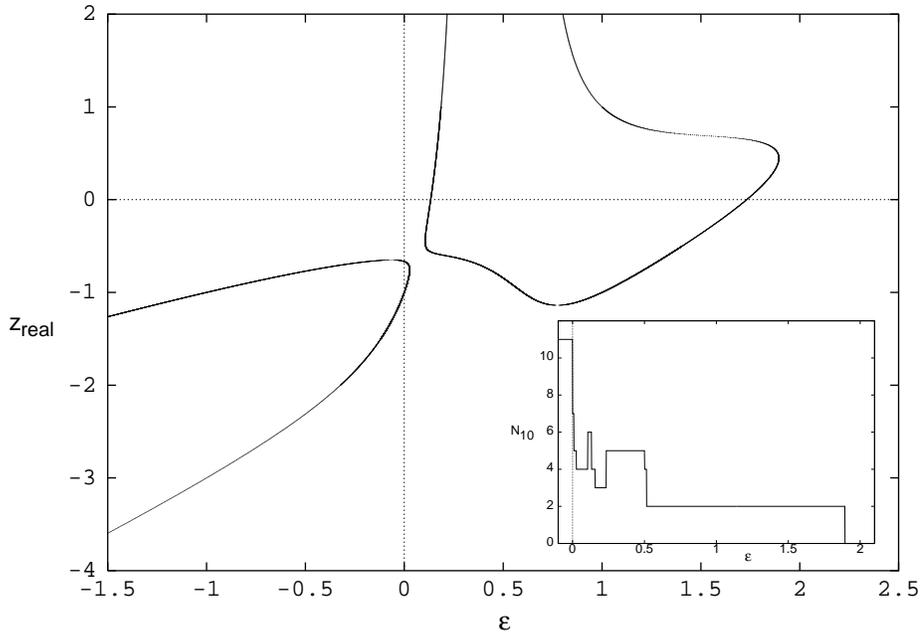}
\caption{The real roots $z$ of $Q_{10}$, as a function of $\epsilon$,
in the  interval
 $[-1.5\, , \, 2.5]$.
We have included the plot of the total number of fixed points 
of $k_{\epsilon}^{10}$ (``P-type'', ``Q-type'' and ``R-type'')
as a function of $\, \epsilon$ in an interval
around $[0\, , \, 2]$.
\label{f:figQ10}
}
\end{figure}
Let us give, for order $10$, a few examples 
of these algebraic ``threshold''
real values of $\, \epsilon $ corresponding
to the real roots of such ``$Q$-type'' polynomials (\ref{Qtype}) :
\begin{eqnarray}
\label{Q10}
&&5 \, \epsilon^2\, -10 \, \epsilon +1 \, \, = \, \, \, 0\, , \,
\, \qquad \qquad 
5 \, \epsilon^4\, -96 \, \epsilon^3 \, + 114 \, \epsilon^2
-40 \, \epsilon\, + \, 1 \, \, = \,  \,\, \, 0 
\end{eqnarray}
The roots of the first polynomial are of the form
(\ref{previous}). The  ``threshold''
 values of $\, \epsilon $ are thus
given by two real roots $\, \epsilon \, \simeq \, \,
.1055 \, $ and $\, 1.8944$.
The second polynomial only 
gives two real roots $\, \epsilon \, \simeq \, .02703 \, $ and $\,\simeq \,
17.9549 $. 

Similar calculations can be performed for the fixed
points of the ``$ P$-type''(see~\cite{Alg})
 corresponding to the line $\, y \, = \,
-z$ (see~\cite{Lyapou}). One can also get
the real roots $z$ of $ \, P_{10}$ as a function of $\epsilon$.
The algebraic values of $\, \epsilon$, occurring in this case, are, 
again, the two roots of the first polynomial
in (\ref{Q10}) together with the
 only two real roots $ \, \epsilon  \, \simeq .1561   \, $
and  $  \,  .5013   \,  $ of
polynomial :
\begin{eqnarray}
\label{P10}
\epsilon^8 \,-26 \, \epsilon^7 \,+343 \, \epsilon^6\, -2052 \,
\epsilon^5 \,+6367 \, \epsilon^4 \,-7178 \, \epsilon^3 \,+3625\,
\epsilon^2\, -824\,\epsilon +64 \, \, = \,
\, \, 0
\end{eqnarray}
as well as two real roots $\, \epsilon\, \simeq .008999  \,  $  
and $\,  \, .1316  \, $
of a  polynomial of degree 24 in $\epsilon$ that will not be
written here.
The last set of points of
 the ``$R$-type'' (see~\cite{Alg}), which corresponds to fixed
points that are neither of the  ``$P$-type'', nor of the 
``$Q$-type'', give 
the following real roots $ \, \epsilon  \, \simeq .2338   $, $\,  \,
.51434   $, and $\, \,  33.2517 $, corresponding to 
polynomial $\, \epsilon^3 \, -34 \cdot \epsilon^2 \,
 +25 \cdot \epsilon \, - 4 \, \,=\, \, \, 0$. On all these algebraic 
values of $\, \epsilon$, one can see a variation
of the total number of fixed points (``P-type'', ``Q-type'' and
``R-type'')
of order $10$ (see insertion in Fig.\ref{f:figQ10}). 
These values are in fact particular
examples of families
of algebraic $\, \epsilon$ values. 
The simplest family of
 singled out {\em algebraic} values
 of $\epsilon$, 
 corresponds to 
the fusion on an $N$-cycle with 
 the $1$-cycle, and reads :
\begin{eqnarray}
\label{previous}
\epsilon \, \, =\, \,  \,
  {{1- \cos (2 \pi \, M / N)}\over{1+ \cos (2 \pi \, M / N)}}\,
  , \, \qquad \quad \hbox{or equivalently :} \qquad \quad
\cos (2 \pi \, M / N) \, = \, \, {{ 1\, -\epsilon } \over { 1\, +\epsilon }}
\end{eqnarray}
 for any integer $N$ (with $\, 1 < M < N/2$, 
$M$ not a divisor of $N$).
Other cycle-fusion mechanisms take place yielding  new 
families of  algebraic values for $\, \epsilon$. For instance the 
coalescence of the $(3\times N)$-cycles in the $3$-cycle,
and the coalescence of the $\, (4\times  N)$-cycles in the $4$-cycle,
yield respectively (with some constraints on the integer $M$ that will
not be detailed here) :
\begin{eqnarray}
\label{moreprevious}
\cos (2 \pi \, M / N) \, = \, \, 
1\, - \, {{ 3} \over {4}} \, {{  \epsilon \, (\epsilon-3)^2 }
 \over { (1-\epsilon) \,  (1+\epsilon)}}\, , \,
\, \qquad 
\cos (2 \pi \, M / N) \, = \, \, 
1\, -32 \, {{ \epsilon \, (1-\epsilon)^2} 
\over {(1+\epsilon)^2 \,  (1-2 \, \epsilon)}}
\end{eqnarray}

\vskip .3cm
{\bf Status of the fixed points :} The fixed point of $\, k_{\epsilon} $,
which is elliptic for $ \epsilon > 0 $, becomes
{\em hyperbolic} for  $ \epsilon < 0 $. For
 three iterations ($N\, = \, 3$) one 
finds that, moving through the  $\, \epsilon\, = \, 1/3\, $ value,
also changes the status of these fixed points 
from elliptic to hyperbolic.
In fact, the algebraic values, like the ones
 depicted in (\ref{previous}) or
 (\ref{moreprevious}), {\em also occur} in such 
changes of  status from elliptic to hyperbolic (see~\cite{Lyapou}).
Therefore, the number of elliptic
fixed points, or hyperbolic fixed points,
is not as ``universal'' as the 
total number of (complex) fixed points, however it has 
``some universality'' : for a given value of $\, N$, the  number of 
 elliptic (resp. hyperbolic) fixed points
depends on $\epsilon$ also by {\em intervals} (staircase function). This 
has, again, to be compared with the
dependence of the growth complexity $\, \lambda$, seen 
as a function of  $\epsilon$,
depicted in Fig.\ref{f:fig1}. The fact that the number
of hyperbolic versus elliptic fixed points, as well as the number
of {\em real versus non-real} fixed points, is modified 
when $\epsilon$ goes through the {\em same set}
of values, like (\ref{previous}) or (\ref{moreprevious}), seems to 
indicate that a modification of the number
of {\em real fixed points is not independent of the 
actual status of these points} (hyperbolic versus elliptic). This 
phenomenon, in fact, corresponds to a quite involved, and interesting,
{\em structure}~\cite{Sulli} that will be sketched in~\cite{Lyapou}.

\subsection{Some expansions for the ``real dynamical zeta function''
and $\, H^{real}\, $.}
\label{Bsome}
Let us recall some results corresponding to
 $\, \epsilon \, = \, .52 $, 
in particular  the product decomposition (\ref{Weil}) of the
 dynamical zeta function~\cite{zeta,McGuire}.
In~\cite{zeta,McGuire}, the number of
 irreducible cycles $N_i$ (see (\ref{Weil})),
as well as the number of irreducible cycles
corresponding to hyperbolic points, elliptic points, real 
points is detailed (see Table I in~\cite{McGuire}).
These results (and further calculations) enable 
to write, for $\, \epsilon \, = \, .52 $,
  the ``real dynamical function'' 
$\, \zeta_{\epsilon}^{real}(t)$ as the following product :
\begin{eqnarray}
\label{eps52}
&&\zeta_{52/100}^{real}(t) \, \, =\, \,  \, 
 {\frac {1}{\left (1-t\right )\left (1-{t}^{3}\right )\, (1-t^{4}
 )\left (1-{t}^{5}\right )^{2}\left (1-{t}^{7}\right )^{2}\left 
(1-{t}^{8}\right )\, (1-t^9)^4 \,  (1-t^{10} )^2}} \\
&&\qquad \qquad \qquad \quad
\,\times \, {{1} \over {(1-t^{11} )^{6} \,
(1-{t}^{12})^{12}
\, (1-{t}^{13})^{16}
}} \, \cdots  \nonumber
\end{eqnarray}
yielding the following expansion for
$\, \zeta_{\epsilon}^{real}(t) \,$ and  $\, H_{\epsilon}^{real}(t)\, $ :
\begin{eqnarray}
\label{eps52expand}
&&\zeta_{52/100}^{real}(t) \, \, =\, \, 
1+t+{t}^{2}+2\,{t}^{3}+3\,{t}^{4}+5\,{t}^{5}+6\,{t}^{6}+9\,{t}^{7}+13
\,{t}^{8}+20\,{t}^{9}+28\,{t}^{10}+40\,{t}^{11}
+65\,{t}^{12}+97\,{t}^{13}\, + \ldots \nonumber \\
&&H_{52/100}^{real}(t) \, \, =\, \,  \,t+{t}^{2}+4\,{t}^{3}
+5\,{t}^{4}+11\,{t}^{5}+4\,{t}^{6}+15\,{t}^{7}+13
\,{t}^{8}+40\,{t}^{9}+31\,{t}^{10}+67\,{t}^{11}+152\,{t}^{12}+209\,{t}^{13} \, + \ldots 
\end{eqnarray}
The number of {\em real} $n$-th cycles of the $P$-type, $Q$-type and
$R$-type, denoted $P_n$, $Q_n$, and $R_n$ respectively, are 
given in Table~\ref{latable} in Appendix~A.
For the $\, R_n$'s one cannot reduce, in contrast with the $P$-type or
 $Q$-type
analysis, the calculations 
to a only one variable : one is obliged to perform 
a first resultant calculation where one eliminates 
one of the two variables 
and another resultant calculation where one eliminates 
the other one, and check back, in the cartesian product of these
possible solutions, the solutions which are actually 
fixed points. In order to get integer 
values that can be trusted, 
 one needs to
perform these (maple) calculations with more than 2000 digits for 
order twelve, but then one faces severe memory limitations
 in the formal
calculations.
We have been able to find integer values for the $R_n$'s for 
orders larger than twelve, however it is clear that these integers
are just lower bounds of the true integers
(not enough precision does not enable to
 discriminate between
 fixed points that are very close). Therefore we prefer not to give
these integers here, and put a ``star'' in Table~\ref{latable}, 
as well as in the forthcoming tables given in Appendix~A, 
when we encounter these computer limitations.

The total number, $\, T_n$, of  {\em real} 
cycles of the $P$-type, $\, R$-type and $Q$-type
actually corresponds to the exponents in the product decomposition
(\ref{eps52}) for the ``real dynamical zeta function''.
Unfortunately, these series are not large enough to ``guess'' any possible 
(and simple, like (\ref{conjec})) rational expression 
for $\, \zeta^{real}(t)$, if any ... Series 
(\ref{eps52expand}), however,  give 
a first ``rough estimate''
for the ``real topological complexity'' $h_{real}$:
 $\, h_{real}\, \,  \,  \simeq \, (97)^{1/13} \, \simeq \,1.4217\, $
or may be  $ h_{real} \, \, \simeq \,   \,209^{1/13}  \, \simeq  \,
1.508 $, {\em clearly smaller} than  the exact algebraic
value for $\, h$ corresponding to 
(\ref{conjec}) :  $\,  h\, \simeq \, 1.61803 $.

Let us consider other values of $\, \epsilon$.  

$ \bullet  $ For $\, \epsilon < 0$,
one finds out that {\em all the fixed points 
seem to be real} and, thus, one can conjecture for $\, \epsilon < 0$
(but $\, \epsilon \, \ne \, -1$) : 
\begin{eqnarray}
\label{zetarealepsneg}
\zeta_{\epsilon < 0}^{real}(t) \, \, =\,
 \, {{ 1-t^2} \over {1-t-t^2}}\, , \qquad \qquad
\hbox{and :} \qquad  \qquad  \,  h_{real} \, = \, h \, \simeq \, \, 1.61803
\end{eqnarray}
The number of cycles of the $P$-type, $Q$-type and $R$-type 
is given order by order in Table~\ref{latable2} in Appendix~A.
These successive integer values for the total number of 
irreducible  real cycles, $\, T_n$,
yield :
\begin{eqnarray}
\label{zetarealepsneg2}
&&\zeta_{\epsilon <0}^{real}(t) \, \, =\,
 \,{\frac {1}{\left (1-t\right )
\left (1-{t}^{3}\right )\left (1-{t}^{4}\right )
\left (1-{t}^{5}\right )^{2}\left (1-{t}^{6}\right )^{2}
\left (1-{t}^{7}\right )
^{4}\left (1-{t}^{8}\right )^{5}\left (1-{t}^{9}\right )^{8}\left (1-{t}^{10}
\right )^{11}}}  \\
&&\qquad \qquad \times 
{\frac {1}{\left (1-{t}^{11}\right )^{18}
\left (1-{t}^{12}\right )^{25}\left (1-
{t}^{13}\right )^{40}\left (1-{t}^{14}\right )^{58}
\left (1-{t}^{15}\right )^{90
}\left (1-{t}^{16}\right )^{135}\left (1-{t}^{17}\right )^{210}
\left (1-{t}^{18}
\right )^{316}}}\, \cdots \nonumber
\end{eqnarray}

 $ \bullet  $ For  $\, \epsilon \, = \, 3 $, 
one has, at every order of iteration, up to order twelve,
a  {\em only one real} fixed point 
(the fixed point of order one but, of course,
 many complex fixed points)  yielding :
\begin{eqnarray}
\label{eq15}
H_{3}^{real}(t) \, \, =\, \, {{t} \over {1-t}} \qquad  \qquad 
\hbox{and :} \qquad \qquad
\zeta_{3}^{real}(t) \, = \, \, {{1} \over {1-t}}\,
\end{eqnarray}
and, for  $\, \epsilon \, \, $ very close to 
 $\,  3$, the expressions of $\, H^{real}(t) \,$
and $\, \zeta^{real}(t) \,$ cannot be distinguished (at the orders
 where we have been able to
perform these fixed points calculations) from :
\begin{eqnarray}
\label{eq16}
H_{\epsilon \, \simeq \, 3}^{real}(t) \, \, \simeq \, \, {{t \cdot (1+t+4 \, t^2)} \over {1-t^3}}
\qquad \qquad 
\hbox{and :} \qquad \qquad
\zeta_{\epsilon \, \simeq \, 3}^{real}(t) \, \simeq  \, \, {{1} \over {(1-t) \cdot (1-t^3)}}\,
\end{eqnarray}
which just correspond to add an additional  3-cycle.

Expressions (\ref{eq15}) are in agreement with the phase portrait of 
Fig.\ref{f:fig10} for $\, \epsilon\, = \, 3$. This
 indicates that, seen as a mapping of two
{\em real variables}, the mapping ``looks like'' an integrable mapping:
the ``real topological complexity''
 $\, h_{real}\, $ seems to be exactly equal
to $\, 1$ for $\, \epsilon \, = \, 3$ (and $\,  h_{real} \simeq 1\, $ for 
$\, \epsilon \,\simeq  \, 3$). The ``real topological entropy'' $\,
\log(h_{real})\, $
seems to be exactly zero for $\, \epsilon \, =\,  3\, $ and
is, thus, drastically different from the
 generic  ``usual'' topological entropy
$\, \log(1.61803 \cdots)\, $. The possible
 foliation of the two-dimensional
space in 
(transcendental) curves is discussed\footnote{
In particular it is shown that, at least, three of the (real) curves
of the phase portrait correspond to {\em divergent series} 
satisfying {\em an exact 
functional equation}~\cite{Enumerativ}.}
elsewhere~\cite{Enumerativ}.
For other values of $\epsilon\, $ the series are not large
enough to ``guess'' a rational expression (if any ...)
for the ``real dynamical zeta function'' $\, \zeta^{real}(t) \,$.

 $ \bullet  $ Miscellaneous examples are given in Appendix~B.
In particular the number of cycles of the $P$-type, $Q$-type and $R$-type 
is given in Table~\ref{latable311} for  $\, \epsilon\, =\, 11/10$,
yielding the following expansion for the real dynamical zeta function :
\begin{eqnarray}
\zeta^{real}_{11/10}(t) \,\, = \,\, \,
1+t+{t}^{2}+2\,{t}^{3}+2\,{t}^{4}+2\,{t}^{5}
+3\,{t}^{6}+5\,{t}^{7}+5\,{t}^{8}+6\,{t}^
{9}+10\,{t}^{10}+12\,{t}^{11}+13\,{t}^{12} +\, \ldots 
\end{eqnarray}
clearly yielding a value for $\, h_{real} \,$ close to one
(may be  $\, h_{real} \, \simeq \,  (13)^{1/12} \, \simeq \,  1.238$)
significantly smaller than $\, h \, \simeq 1.618\,$.
This result has to be compared with the
equivalent one for  $\epsilon \, = \, 9/10$ or 
for  $\epsilon \, = \, 21/25\, $ :
\begin{eqnarray}
\label{21sur25}
&&\zeta^{real}_{21/25}(t) \,\, = \,\, \,
{\frac {1}{\left (1-t\right )\left (1-{t}^{3}\right )\left (1-{t}^{4}
\right )\left (1-{t}^{7}\right )^2\left (1-{
t}^{8}\right )\left (1-{t}^{10}\right )^{2}
\left (1-{t}^{11}\right )^{4}\left (1-{t}^{12}\right )^{2} }}\,  \cdots
\,  \\
&& \qquad \qquad \,\, = \,\, \,
1+t+{t}^{2}+2\,{t}^{3}+3\,{t}^{4}+3\,{t}^{5}+4\,{t}^{6}+7\,{t}^{7}
+9\,{t}^{8}+10\,{t}^{9}+15\,{t}^{10}+23\,{t}^{11} +28\,{t}^{12}+ \,\, \ldots \nonumber
\end{eqnarray}
yielding a larger value for $\, h_{real} \,$ :
 $\, h_{real} \, \simeq \,  (28)^{1/12} \, \simeq \,  1.32$.
This expansion is actually compatible with the following simple
 rational expression and for its logarithmic derivative $\, H_{21/25}^{real}$ :
\begin{eqnarray}
\label{compat21sur25}
\zeta^{real}_{21/25}(t) \,\, = \,\, \,
{\frac {1+{t}^{2}}{1-t+{t}^{2}-2\,{t}^{3}}}\qquad \hbox{and :} \qquad 
 H_{21/25}^{real}  \,\, = \,\, \,
{\frac {t\left (5\,{t}^{2}+2\,{t}^{4}+1\right )}{\left
(1+{t}^{2}\right )\cdot \left (1-t+{t}^{2}-2\,{t}^{3}\right )}}
\end{eqnarray}
Note that all the coefficients of the expansion of
the rational expression (\ref{compat21sur25})
and of its logarithmic derivative $\, H_{21/25}^{real}$
 are positive (in contrast with  
Pade approximation (\ref{simpledeuxsurtrois}) given in Appendix C for $\, \epsilon \, =
\, 2/3$ which is ruled out because coefficient $\, t^{54}$ of its
expansion is negative). If this simple rational expression 
is actually the exact expression for the real dynamical zeta function
$\, \zeta^{real}_{21/25}(t)  \,$
this would yield the following algebraic value for  $\,
h_{real}, $ : 
$\, h_{real}(21/25)\, \simeq \, 1.353209964$.

For $\, \epsilon \, = \, 9/10$,
one gets the same product decomposition,
at least up to order ten. The number of $n$-th
 cycles of the $P$-type, $Q$-type and
$R$-type for  $\, \epsilon \, = \, 9/10$ are given in Appendix B.
One thus sees that  $\, h_{real} \,$ decreases 
when  $\, \epsilon \,$ crosses the  $\, \epsilon \,= \, 1$
value.

$\bullet$  For $\, \epsilon \, = \, 1/4$ we have obtained (see
Appendix~B) :
\begin{eqnarray}
\label{unsurquatrereel}
&&\zeta^{real}_{1/4}(t) \,\, = \,\, \,
{\frac {1}{\left (1-t\right )\left (1-{t}^{3}\right )\left (1-{t}^{4}
\right )\left (1-{t}^{5}\right )^{2}\left (1-{t}^{7}\right )\left (1-{
t}^{8}\right )\left (1-{t}^{9}\right )^{3}\left (1-{t}^{10}\right )^{2
}\left (1-{t}^{11}\right )^{4}\left (1-{t}^{12}\right )^{4}
\left (1-{t}^{13}\right )^{8}}} \,\cdots
\,  \nonumber \\
&&\qquad  \,\, = \,\, \, 1+t+{t}^{2}+2\,{t}^{3}+3\,{t}^{4}+5\,{t}^{5}
+6\,{t}^{6}+8\,{t}^{7}+12\,{t}
^{8}+18\,{t}^{9}+25\,{t}^{10}+34\,{t}^{11}\,
 +48\,{t}^{12}\, + \, 70 \,{t}^{13}\, + \, \ldots 
\end{eqnarray}
The ``non-generic'' values $\epsilon \, = \, 1/m\, $ and $\, (m-1)/(m+3)$
require a special and careful analysis\footnote{Some fixed
points near these ``non-generic'' values ($\epsilon \, \simeq \, 1/m$), 
disappear on these very values stricto sensu : they become singular.
One has to verify carefully that all the points, obtained 
in such calculations, are fixed
points and not singular points.}. However, similarly to what was seen
for  $\,\zeta(t) \,$,
 one clearly verifies 
on all these ``real 
dynamical zeta function'' $\,\zeta^{real}(t) \,$
that the coefficients in these expansions are continuous in 
$\, \epsilon\, $ near these points except on these very values 
of $\, \epsilon$ where
one gets smaller integers and thus smaller values for $\, h_{real}$
(the limit on the left 
and on the  right of  $\, h_{real}$ are equal and larger
than $\, h_{real}$ on these very ``non-generic'' values).

The numbers of irreducible {\em real} $n$-cycles of
the $P$-type, $Q$-type and $R$-type 
are given in Appendix B for miscellaneous
values of $\, \epsilon\, $ : $\epsilon = 11/100$, $\epsilon = 5$,
$\epsilon = 10$, $\epsilon = 50$.
We also give, in Appendix C, for  various values of $\, \epsilon \,$
($\epsilon=9/50$,   $\epsilon=31/125$,
 $\, \epsilon \, = \, 12/25$, $\epsilon\, =\, 17/25$, 
   $\epsilon=\, 66/125$,  $\epsilon=\, 3/4$,
 $\epsilon=\, 3/2$),  the product decomposition and
expansions for $\, \zeta^{real}_{\epsilon}(t) \,$ up to order 11.

Similar calculations of the expansions of $\, H_{\epsilon}^{real}(t) \, $
and $\, \zeta_{\epsilon}^{real}(t) $, for many other values 
of the parameter $\, \epsilon$,
have been performed and will be detailed elsewhere. 
All these results confirm that  $\,h_{real}\,  $ {\em varies with} 
$\, \epsilon $ when $\, \epsilon $ is positive, while
 $\, h_{real}\, $ is constant (except $\epsilon = -1$)
when $\epsilon <0$. When $\, \epsilon $ is positive,
the estimates of 
$\, h_{real}\, $ 
 are in agreement with the ``visual complexity''
as seen on the phase portraits (see the previous section).
In particular one finds that
 $\, h_{real}\, $ roughly 
decreases as a function of $\epsilon$ in the  
intervals $[0^{+}\, , \, \simeq 1/10]$ and 
$[\simeq 1/3 \, , \, 1^{-}]$, and increases in the 
interval $[\simeq 1/10 \, , \,\simeq 1/3 ]$,
(with a sharp decrease near $\, \epsilon \simeq 1/2$
and  $\, \epsilon \simeq 1$),
that $\, h_{real}\, $
is close or very close to one when $\, \epsilon\, $ belongs to an
interval $[1^{+}\, , \, \simeq 16]$, that $\, h_{real}\, $
grows monotonically with  $\, \epsilon\, $ 
for  $\, \epsilon\, > 16 $ to reach some asymptotic 
value in the  $\, \epsilon \,\, \rightarrow \,\,
\infty\, $ limit. It will be seen, in the next subsection,
that the  ``real topological
 complexity''  $\,  h_{real}\, $, 
in the $\, \epsilon \,\, \rightarrow \,\,
\infty\, $ limit, tends to
a value  $\,  h_{real}\, \simeq \, 1.429  \, \, $
clearly different, again, from the generic ``topological complexity''
$\, h \, \simeq \, 1.618$.

\subsection{Seeking for rationality for the ``real dynamical zeta
function''.}
\label{seek}

 Recalling the large number of rational
expressions, obtained for the
 dynamical zeta functions~\cite{zeta,McGuire} and the
degree generating functions~\cite{zeta,complex,BoMa95},
one may have a rationality ``prejudice'' for these
``real dynamical zeta functions'' $\, \zeta_{\epsilon}^{real}(t) $,
calculated for a {\em given} value of $\epsilon$.
However, the 
occurrence of any symbolic dynamic, and associated 
Markov partition, is {\em far from being natural in
this real analysis framework}~\cite{Easton,Conley,Fried87}.
If one bets on the rationality of  the
real dynamical zeta function $\, \zeta_{\epsilon}^{real}(t) \,$
(see (\ref{defreal})),  it must, however, be clear that
 $\, \zeta_{\epsilon}^{real}(t) $ 
depends on $\, \epsilon \, $ in a very complicated way (piecewise
continuous
functions, devil's
staircase ? ...).  If, for some given
 value of the parameter $\, \epsilon$, the
partial dynamical zeta function
 $\, \zeta_{\epsilon}^{real}(t) $ actually corresponds to a rational  expression,
one should, in fact, have an {\em infinite set}
 of such rational expressions
associated with the infinite number
of steps (intervals\footnote{In contrast
 with the situation for the ``customary'' dynamical zeta
function
which is equal to one generic universal
 expression (like (\ref{conjec})),
up to a (zero measure) set of values of
 $\epsilon$ (see Fig.\ref{f:fig1}).} 
in $\epsilon$) in the ``devil's staircase''. The actual 
location of these  ``steps'', that is, the 
limits of these intervals in $\epsilon$,
corresponds to an infinite number of values 
of $\epsilon$ like (\ref{previous}) or (\ref{moreprevious}) 
(and others~\cite{Lyapou}). 
For a given $\epsilon$, the calculations of the first terms of
the expansion of the ``real dynamical 
zeta function''  $\, \zeta_{\epsilon}^{real}(t) \,$ do not
rule out, at the order for which we have been able
to perform these calculations (ten, eleven),
{\em rational expressions} (see for instance~\cite{Enumerativ}).

The number  of  {\em real} $n$-th cycles of the $P$-type, $Q$-type and
$R$-type, for $\epsilon \, = \, 50$, is given in 
Table~\ref{latable32} in Appendix~B.
One remarks that, at order eleven, the number 
of irreducible {\em real } cycles,
and therefore the expansion of the ``real dynamical zeta function''
are the same for 
$\epsilon\, = \, 50\, , \, 100\, , \, 1000\, , \, \ldots$
For  $\epsilon=50$,  $\epsilon=100$ one has a product 
expansion for the dynamical zeta function identical,
up to order eleven,  to 
the product  expansion corresponding
 to the  $\epsilon\, $ large limit (see
 (\ref{expanrealArnold1inf}) below). These expansions 
 are, however, different at order twelve, see Appendix B.

\vskip .3cm

$\bullet$ {\bf $\epsilon $ large.} For $\, \epsilon \, $ large enough
 one gets the following cycle product decomposition :
\begin{equation}
\label{expanrealArnold1inf}
\zeta^{real}_{\epsilon = \, \infty}(t) \, = \, \, \,  \, \, 
{{1} \over { (1-t) \, (1-{t}^3) \, (1-{t}^{5})^{2}
\, (1-{t}^7 )^{2}\left (1-{t}^{8}\right )^2 \, (1-{t}^{9} )^{
2}\, (1-{t}^{10} )^{3}\, (1-{t}^{11} )^{4}
 \, (1-{t}^{12})^6 }} \cdots
\end{equation}
corresponding to the  number of {\em real} $P$-type, $Q$-type and
$R$-type $\, n\, $ cycles for $\, \epsilon \, =  \, 20000$
given in Table~\ref{latable3} in Appendix B. 
Note that one gets the same table (up to order 16) 
for $\, \epsilon \, = \, 1000, \, 100000\, , \, 1000000$.

One finds out easily that these results, for
the ``real dynamical zeta function'' $\, \zeta^{real}(t) $,
 are (up to order twelve)  actually 
in {\em perfect agreement} with
 (the expansion of) the rational expression :
\begin{eqnarray}
\label{zetarealinfty}
\zeta^{real}_{\epsilon = \, \infty}(t) \, \, \, = \, \, \, \,
 {{ 1+t} \over {1-t^2-t^3-t^5}}\, \, \, = \, \,  \, \,
 {{ 1-t^2} \over {(1-t-t^2) \, + t^4 \cdot (1-t+t^2)}}
\end{eqnarray}
yielding an algebraic value for  $\, h_{real}\, $ :
 $\, h_{real}\,\, \simeq \, 1.4291  $. 
If one ``believes'' in some symbolic dynamic coding interpretation,
or in the existence of a linear transfer operator\footnote{For
more details on  linear transfer
operators in a symbolic
dynamics
framework, see for instance~\cite{Ru78,Bo73,Bo75,Bo70}.}, 
matrix $\, \, A$, such that the denominator of (\ref{zetarealinfty}),
$  \, \,1 -t^2 -t^3 -t^5 $, can be written as 
$\, \det(Id-t\cdot A)$, one finds that a possible choice
for this transition matrix is :
\begin{eqnarray}
\label{Markovtransi}
A \, = \, \, 
\left [\begin {array}{ccccc}
 0&0&1&0&1\\
\noalign{\medskip}1&0&0&1&0\\
\noalign{\medskip}
1&0&0&0&0\\
\noalign{\medskip}0&0&1&0&0\\
\noalign{\medskip}0&1&0&0&0
\end {array}\right ]
\end{eqnarray}
In contrast with a Markov's transition matrix
the previous matrix is not such 
that the sums of the entries in each row,
or column, are equal.

\section{``Real Arnold complexity''}
\label{RAC}
Let us recall, again, the identification between $\, h$, 
the (exponential of the) 
topological entropy, and $\, \lambda $,
the (asymptotic of the) 
Arnold complexity~\cite{zeta}. Similarly to the topological entropy,
the Arnold complexity can be ``adapted'' to define 
a ``{\em real} Arnold complexity''. 
The Arnold
 complexity counts the number
of intersections between a fixed (complex projective) line
and its $N$-th iterate~\cite{A} : let us now count, here,  the number
of {\em real} points which are the intersections
 between a {\em real fixed line} and 
 its $N$-th iterate. With this restriction to {\em real} points
we have lost
``most of the universality properties'' of the ``usual'' 
(complex) Arnold complexity.
For various values of $\, \epsilon $, we have calculated the number
of  intersections of various (real) lines with their $N$-th
iterates. In contrast with the ``usual''  Arnold complexity~\cite{A},
which does not depend on the (complex) line one iterates (topological
invariance~\cite{S67}), it is clear that the
 number of real intersections depends on the chosen line,
but one can expect that the {\em asymptotic behavior} of these 
numbers for $\, N$ large enough, will not  depend too much of the actual
choice of the (real) line one iterates. 
Actually, we have discovered on this very example, that this seems to
be the case (except for some non generic lines).
Furthermore, the real line $\, y \, = \, (1-\epsilon)/2 $, which is known
to play a particular role  for mapping (\ref{yz}) 
(see section (\ref{Creal})),
is very well-suited to perform these numbers of intersections
calculations : for this particular line the successive numbers of
intersections are extremely regular, thus enabling 
to better estimate this asymptotic behavior $\, \lambda_{real}^N\, $
of the  ``real Arnold complexity'', but of course
 similar calculations can be performed with an arbitrary (generic) line.
$\lambda_{real}\, $ can be seen as the equivalent, for {\em real mappings},  of the
 growth complexity $\, \lambda\, $
 (see section (\ref{seminum}) and~\cite{complex}). 
Let us try to get $ \lambda_{real}\, $ for various values of $\, \epsilon$.

\begin{figure}
\psfig{file=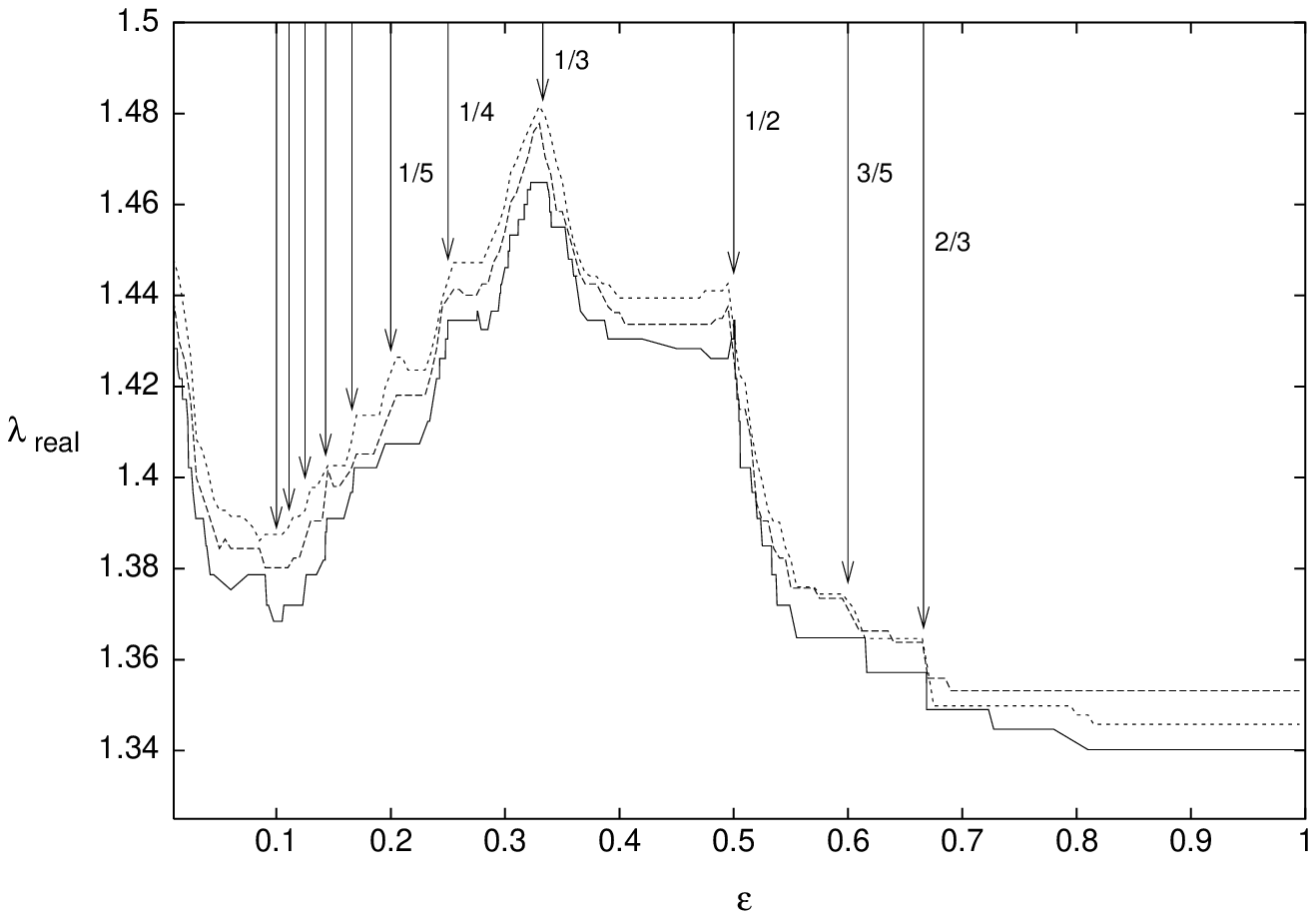}
\caption{A representation of $\, \lambda_{real}(\epsilon)\,\ $ 
by $\, {\cal A}_{N}^{1/N} $,
 as a function of $\, \epsilon$, in the $\, [0\, , \, 1]\, $ interval, for $\, N\, = \, 13$
(full line)
$14$ (dashed  line),  and $ 15$ (dotted line). The 
singled out $\epsilon \, = \, 1/m$ values, 
 for $\, m\, = \,  \,  2\, , \, 3\, , \,  4\,, \,  5\, , \, \ldots
$, and  $\, \epsilon \, = \, (m-1)/(m+3)\, $
for $\, m\, = \, 7\,, \,  9\, , \, , \, \ldots\, $
clearly play a special role for  these various ``staircase
functions'',
in the  large  $N$ limit.
\label{f:fig11}
}
\end{figure}

Similarly to the semi-numerical method
 detailed in section (\ref{seminum}),
we have developed a C-program using again 
the multiprecision library
gmp~\cite{C}, counting the number of (real) intersections of the 
$\, y \, = \, (1-\epsilon)/2\, $ real line
with its $N$-th iterate. This program 
does not calculate the precise location of the intersection points :
it is based on the {\em Sturm's theorem}\footnote{Assuming that 
a polynomial $\, P(x)\, $ has no multiple roots,
one can build a finite series of polynomials corresponding
to the successive Euclidean division of $\, P(x)\, $ by 
its first derivative $\, P'(x)$. See for instance~\cite{Ma66} for more
details on the Sturm sequences and Sturm's theorem.
}. All these calculations
have been cross-checked by a (maple) program calculating 
these numbers of intersections using the sturm procedure in
maple\footnote{The sturm procedure one can find in maple gives
the number of real roots of a polynomial in any interval [a,b],
 even  the interval $]\,  -\infty\, , \,  \,+\infty\,  [$. 
The procedure sturm uses {\em Sturm's theorem} to
 return the number of real roots
of polynomial P  in the interval [a,b]. The first argument of this
sturm procedure is a {\em Sturm sequence} for P, which can be obtained
with another  procedure, the procedure 
sturmseq which returns the Sturm sequence as a list of polynomials and replaces multiple
  roots by single roots.}.
The results of these calculations are given in Fig.\ref{f:fig11}. 

Let us denote by $\, {\cal A}_N\, $ the number of  (real) 
intersections for the $N$-th iterate.
In order to estimate  ``real growth complexity'' 
$\, \lambda_{real} \, $ we have plotted 
$\, {\cal A}_N^{1/N}$, for various values of the number of iterations 
($ N \, = \, 13  ,  \, 14  , \, 15 $), 
as a function of $\, \epsilon$, in the range $[0\, , \, 1]$ where 
$\, \lambda_{real}\, $ has a quite ``rich'' behavior.
This behavior should 
be compared with the ``universal'' behavior of Fig. 1.
On Fig.\ref{f:fig11}, one remarks that the singled out values $\,\epsilon \,
= \, 1/m$, as well as  $\, \epsilon \, = \, (m-1)/(m+3)\, $
for $\, m\, = \, 7\,, \,  9\, , \, , \, \ldots $, 
 seem, again, to play a special role in the large $N$
limit. Of course, recalling the results of section (\ref{Creal}),
 it is clear that  $\, {\cal A}_{N}^{1/N}\, $
is a staircase function of $\, \epsilon  $, for $N$ finite, the limits
of each interval corresponding to
 {\em algebraic values} (like (\ref{previous}))
sketched in sections (\ref{Bsome}), (\ref{Creal}). These algebraic
values form an {\em infinite set} of values which accumulate everywhere 
in the $\, [0, 1]$ interval.
What is the limit of these functions $\, {\cal A}_N^{1/N}(\epsilon)\, $
when $\, N$ gets large : a devil's
staircase or a (piecewise) continuous function?  The ``shape''
 of $\, {\cal A}_{N}^{1/N} $, as a function of 
$\,\epsilon $, is ``monotonic enough'' (see Fig.\ref{f:fig12} below) 
in different intervals, namely 
$ \epsilon\,< 0$ , and   
 in the intervals of $\, \epsilon\,$ roughly given by :
 $\, [0^{+}\, , \simeq 1/10]$, 
 $\,[\simeq 1/10 \,,\, \simeq 1/3]$,  $\,[\simeq \,1/3\, , \,  1^{-}]$, 
$\, [1^{+}\, , \, \simeq 16.8]$ and $\, [ \simeq 16.8\, , \, \infty]$,
that one may expect that the infinite accumulation 
of these  algebraic values (like (\ref{previous})) could yield a 
perfectly continuous function $\, \lambda_{real}(\epsilon)$ 
(except on the non-generic values 
$\, \epsilon \, = \, 1/m$ and $\epsilon \, = \, (m-1)/(m+3)$)
and not a devil's staircase-like function.
This question remains open at the present moment. 
When $\epsilon$ varies from $\, -\infty \, $ to $\, \infty\, $ 
the behavior of the ``real growth complexity'' $\, \lambda_{real} $, as a function of 
$\,\epsilon$, is not as ``rich'' as in the interval 
$[0^{+}\, , \, 1^{-}]$ depicted in Fig.\ref{f:fig11}. One finds that 
 $\, \lambda_{real}\, $ is close, or extremely close to $\, 1$, in a quite large 
interval $\, [1^{+}\, , \, \simeq 16.8]$ and that it  increases 
monotonically with $\, \epsilon$ in the  $\, [ \simeq 16.8\,
, \, \infty]$ interval
to reach some asymptotic value in the  $\,\epsilon \, \rightarrow
\, \infty$ limit. In fact, a logarithmic scale in $\, \epsilon$
is better suited to describe  $\, \lambda_{real}\, $ as a function of 
$\,\epsilon\, $. Fig.\ref{f:fig12} represents  $\, \lambda_{real} $, more precisely
$\, {\cal A}_{13}^{1/13}$,
as a function of $\, \log(2+\epsilon)$.  For $\, \epsilon\, = \, -1\, , \,
0\, , \, 1/3\, , \, 1/2\, , \, 1\, $ we know that  $\, \lambda_{real}\, $ 
will be exactly equal to $1$ (integrable
 cases~\cite{BoHaMa97}). On these points 
(represented by  squares on Fig.\ref{f:fig12}), as well as on the 
$\epsilon \, = \, 1/m$ and $\epsilon \, = \, (m-1)/(m+3)$
non-generic points, $\, \lambda_{real} \, $ is {\em not continuous}
 as a function of $\, \epsilon$. We have not represented these other non
generic points. They have to be calculated separately. 
\begin{figure}
\psfig{file=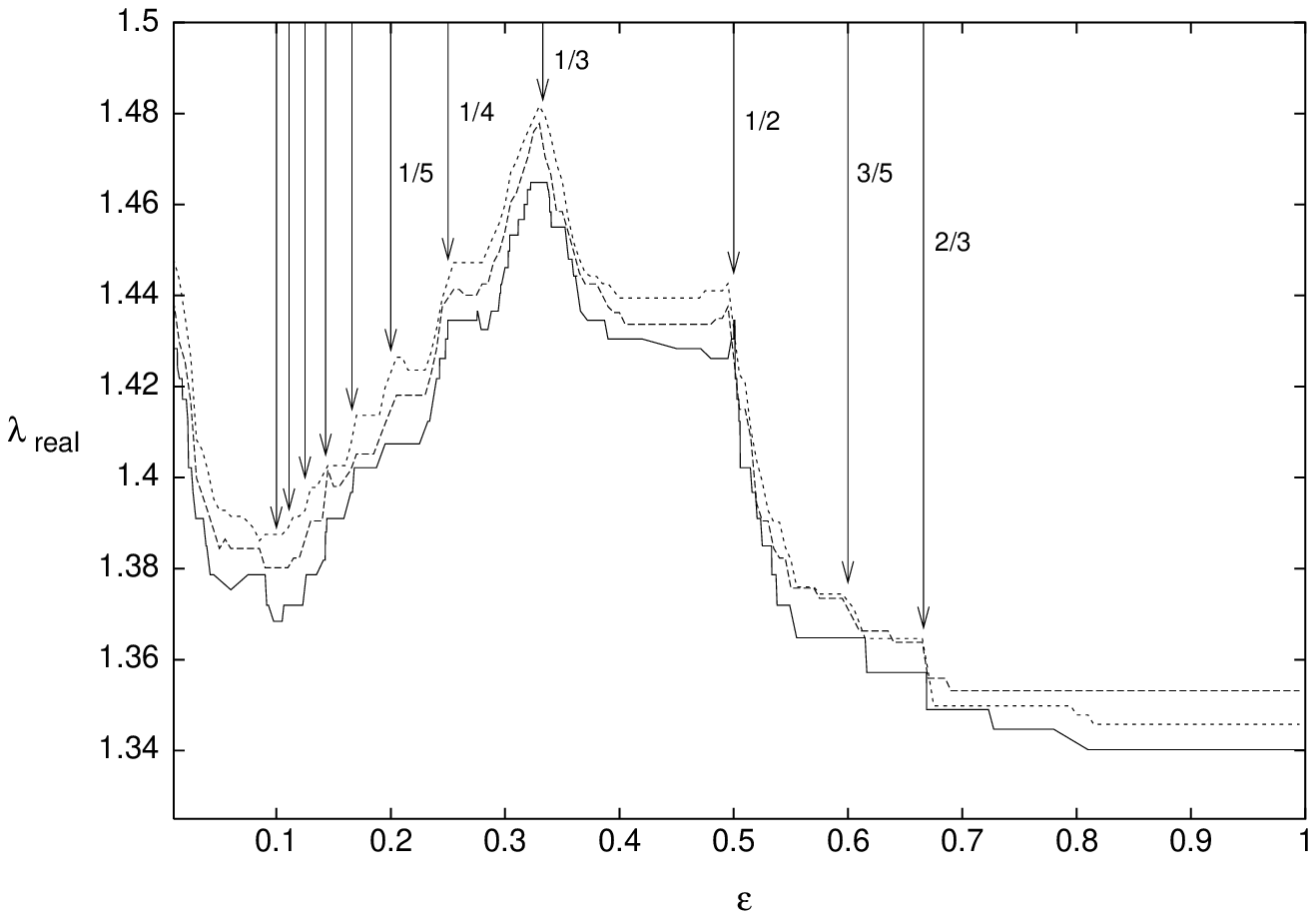}
\caption{A representation of $\, \lambda_{real} \,\ $ 
by $\, {\cal A}_{13}^{1/13} $, as a function of $\, \epsilon
 $, in a  $\, \, \log(2+\epsilon)$ logarithmic scale. The integrable
points  $\, \epsilon\, = \, -1\, , \,
0\,, \, 1/3\, , \, 1/2\, , \, 1\, $ are represented by a square.
\label{f:fig12}
}
\end{figure}
A first estimate of  $\, \lambda_{real} $, for
$\epsilon$ large, is $ \,\lambda_{real}\, \simeq  \,  \, (214)^{1/15} \, \simeq \, 
1.43008  \, $. We are now ready 
to compare the ``real topological entropy'' 
and the ``real Arnold complexity'' for different values 
of $\,\epsilon $, and see if the identification,
 between $\, h $, (the exponential
of)  the topological entropy, 
and $\, \lambda $, characterizing the (asymptotic behavior 
of the) usual Arnold complexity,
also holds for their ``real partners'' namely $\, h_{real}\,$
and $\, \lambda_{real}$. Actually, one finds that this identification 
(which is obviously true for $\, \epsilon \, < \, 0$)
also holds for $\, \epsilon \, = \, 3$ and give numerical results for
various values of $\, \epsilon \, $ (for which we have estimated
the ``real'' topological entropy  $\, \log(h_{real})\,$
 (see section (\ref{Bsome}))),  
{\em quite compatible with this
identification}. In particular, for $\epsilon $ large, we do see
that these two ``real complexities'' {\em give extremely close results},
namely
$\, h_{real} \simeq 1.4291  \,\, $ versus  $ \,\,\lambda_{real}\, \simeq \, 1.43$.

\section{ ``Real Arnold complexity'' generating
 functions : seeking for rationality.}
\label{RACsection}

Similarly to the introduction
of the ``real dynamical zeta functions'' $\, \zeta^{real}(t) $,
one can introduce the generating function 
of the previous ``real Arnold complexities'' $\,  {\cal A}_N \, $  :
\begin{eqnarray}
\label{defrealArnold}
{\cal A}_{\epsilon}(t) \, = \, \,
\, \, \sum_N {\cal A}_N \cdot t^N
\end{eqnarray}
 Recalling the large number of rational
expressions, obtained for the
 dynamical zeta functions~\cite{zeta,McGuire} and the
degree generating functions~\cite{zeta,complex,BoMa95},
one may have, again, a rationality ``prejudice'' for these
``real Arnold complexity generating  functions''. Let us 
try to see if the expansions of these
generating functions
 $\, {\cal A}_{\epsilon}(t) \,$ could coincide,
for some given values of $\, \epsilon$,
with the expansion of some
(hopefully simple) rational expressions.

In order to compare
more carefully $\, h_{real}\, $ and $\, \lambda_{real}$,
and find some possible non trivial rational expressions
for $\, {\cal A}_{\epsilon}(t)  $, let us give, in 
the following, miscellaneous  expansions of
$\, {\cal A}_{\epsilon}(t) \,$ for various values of $\, \epsilon$.

\subsection{Expansions for the ``real Arnold 
complexity'' generating functions}
\label{expanRAC}

In fact, we have not only
calculated the real Arnold complexities $\, {\cal A}_{13}$, 
$\, {\cal A}_{14} $ and  $\, {\cal A}_{15} $
required to plot Fig.\ref{f:fig11} and Fig.\ref{f:fig12},
but actually obtained all the coefficients 
up to order 13 for 2000 values of $\epsilon$,
and up to order 15 for 200 values of $\epsilon$.
We thus  have the expansion of $\, {\cal A}_{\epsilon}(t) \,$
up to order 13 (resp. 15) for several thousands
of values of $\, \epsilon$.

Let us first give the expansion of $\, {\cal A}_{\epsilon}(t) \,$
for $\, \epsilon \, = \, .52\, $ for which the 
expansion of the real dynamical zeta function
has been given previously (see (\ref{eps52})
 and (\ref{eps52expand})).
One gets the following  expansion :
\begin{eqnarray}
\label{expanrealArnold52}
 {\cal A}_{52/100}(t) \, = \, \, 
t\, + t^2\, +2\,{t}^{3}+\, 3\, {t}^{4}
+5\,{t}^{5}\, +\, 6\,{t}^{6}\, + 11 \,{t}^{7}
+11 \,{t}^{8}+16 \,{t}^{9}+29\,{t}^{10}+
\,33\, {t}^{11} +\,46\, {t}^{12} +\,73\, {t}^{13} \, +
\ldots 
\end{eqnarray}
This series  yields a first rough approximation of $\, \lambda_{real}\, $
corresponding to $\, \lambda_{real}\, \simeq  \, (73)^{1/13}\, \simeq  \,1.391  $,
clearly smaller compared to the generic complexity $\lambda \simeq
1.61803$, and in  good enough
 agreement with 
the estimation of $\, h_{real}\, $
 one can deduce
from  (\ref{eps52expand}), namely $\, h_{real}\,
\simeq \, \, (93)^{1/13}\,\simeq\,  1.417 $. Of 
course these two series are to short to see
if an identity like  $\, h_{real}\,= \, \lambda_{real}\, $ really holds.

Considering $\, h_{real}\,$ as a function of $\, \epsilon $,
it is clear that the  general shape of this ``curve''
looks extremely similar to the curve corresponding
to $\, \lambda_{real}\, $ as a function of $\, \epsilon \, $ 
(see Fig.\ref{f:fig11} and Fig.\ref{f:fig12}) : 
it is also constant for $\, \epsilon \, $
negative, gets to close to 
1 around $\, \epsilon \, \simeq \, 3$,
grows monotonically for  $\epsilon \, > \, 10$
and tends to a non-trivial asymptotic value 
 $\, h_{real}\,\simeq \, 1.4291$.

Therefore, in order to get some hint on the relevance of a  
possible   $\, h_{real}\,= \, \lambda_{real}\, $ identity,
it is necessary to see if this relation 
holds for various values of $\, \epsilon\, $ for which  $\, h_{real}$,
and $ \, \lambda_{real} $, can be calculated exactly 
or for which very good  approximations of 
 can be obtained, namely $\, \epsilon < \, 0$,
all the integrable values, or $\epsilon \, = \, 3$ and its 
neighborhood ...

For {\em any negative value} of $\, \epsilon$  (except $\epsilon \, = \,
-1$ see below (\ref{expanrealArnold3})) the expansion of
 the ``real Arnold complexity'' generating function 
$\, {\cal A}_{\epsilon }(t) \,$
is equal, up to order 15, to :
\begin{eqnarray}
\label{expanrealArnold9}
{\cal A}_{\epsilon <0}(t) \, = \, \, \, {{t} \over {1-t-t^2}}
\end{eqnarray}
in agreement with (\ref{zetarealepsneg}).

For $\, \epsilon \, = \, 3\, $ the generating function 
for the ``real Arnold complexity'' generating function 
$\, {\cal A}_{\epsilon }(t) \,$ is equal, up to order 15,
to the simple {\em rational} expression:
\begin{eqnarray}
\label{expanrealArnold11}
{\cal A}_{3}(t) \, = \, \, {{t} \over {1-t}}
\end{eqnarray}
which is in perfect agreement with the 
result for the ``real dynamical zeta function''
$\, \zeta^{real}_{\epsilon=3}(t)\, $
(see (\ref{eq15})).
For  $\, \epsilon \, \, $ very close to 
 $\,  3$  one gets :
\begin{eqnarray}
\label{expanrealArnold11sim}
{\cal A}_{\epsilon \simeq 3}(t)
 \, \,\simeq \,\, \, {{t} \over {1-t}}\, +
\,  {{t^3} \over {1-t^3}}\,\, = \, \,
{\frac { t \cdot (1 \,+t\, +2\,{t}^{2})}{1\, -\, t^{3}}}
\end{eqnarray}
again in good agreement with (\ref{eq16}).

\vskip .3cm 
{\bf Integrable values for $\, \epsilon$ and around.} For $\,
\epsilon \, = \, 1/2\, $ 
the generating function 
for the ``real Arnold complexity'', $\, {\cal A}_{\epsilon }(t) $,
 is equal, up to order 38,
to the expansion of the rational expression :
\begin{eqnarray}
\label{expanrealArnold2}
{\cal A}_{1/2}(t) \, = \, \,
{\frac {t\left (1-{t}^{7}\right )}{\left (1-t\right )^{2}\left (1-{t}^{5
}\right )\left (1+t\right )}}\, + \, \, 
{\frac {{t}^{4}\left (1-{t}^{9}\right )}{\left (1-t\right )\left (1-{t}^
{5}\right )\left (1-{t}^{6}\right )\left (1+t\right )}}
\, + \, \, 
{\frac {{t}^{5}}{\left (1-{t}^{5}\right )\left (1-{t}^{3}\right )\left (
1+t\right )}}
\, + \, \, 2\,{\frac {{t}^{28}}{1-{t}^{5}}}
\end{eqnarray}
to be compared with $A_{1/2}(t)$ given in (\ref{corner}).

For $\, \epsilon \, = \, 1/3\, $ the calculations
corresponding to the generating function 
for the ``real Arnold complexity'' are, in contrast, quite trivial
yielding :
\begin{eqnarray}
\label{expanrealArnold2bis}
{\cal A}_{1/3}(t) \, \, \, \, = \, \,  \,
{{ t \cdot (1+t)} \over {1-t^3}} \, \, = \, \,  \,
 \, \,A_{1/3}(t)
\end{eqnarray}

For $\, \epsilon \, = \, -1\, $ the generating function 
for the ``real Arnold complexity'' is equal, up to order 15,
to the rational expression :
\begin{eqnarray}
\label{expanrealArnold3}
{\cal A}_{-1}(t) \, = \, \, {{t} \over {1-t^2}} \, \, = \, \, \, A_{-1}(t)
\end{eqnarray}

For $\, \epsilon \, = \, +1\, $ the generating function 
for the ``real Arnold complexity'' is equal, up to order 15,
to the rational expression :
\begin{eqnarray}
\label{expanrealArnold4}
{\cal A}_{1}(t) \, \, = \, \, 
 \,{{t} \over {(1-t^2) \cdot (1-t)}} \,  \,=  \,\, \, A_{1}(t)
\end{eqnarray}
all these results have to be compared with the generating functions
(\ref{corner}).

\vskip .3cm 
{\bf Non-generic values for $\, \epsilon$ and around.} The
non-generic values 
of $\, \epsilon \, $ require a special
attention. For instance for $\, \epsilon \,=1/4 $
 one obtains the following expansion\footnote{These maple
calculations
have been performed with 6000 digits, but they are already stable with
2000 digits.} :
\begin{eqnarray}
\label{expanrealArnold14}
&&{\cal A}_{1/4}(t) \, = \, \,t+{t}^{2}+2\,{t}^{3}+3\,{t}^{4}+5\,{t}^{5}
+6\,{t}^{6}+8\,{t}^{7}+11\,{t}^{8}+17\,{t}^{9}
+23\,{t}^{10}+31\,{t}^{11} +44\,{t}^{12}+63\,{t}^{13}\nonumber \\
&&\qquad \qquad +90\,{t}^{14}+128\,{t}^{15}+183\,{t}^{16}\, + \ldots \nonumber
\end{eqnarray}
and for  $\, \epsilon \,=1/5 $
one gets  :
\begin{eqnarray}
\label{expanrealArnold17}
&&{\cal A}_{1/5}(t) \, = \,
t+{t}^{2}+2\,{t}^{3}+2\,{t}^{4}+4\,{t}^{5}+4\,{t}^{6}+6\,{t}^{7}+7\,{t}^{8}
+12\,{t}^{9}+15\,{t}^{10}+19\,{t}^{11}+28\,{t}^{12}+33\,{t}^{13}\nonumber \\
&&\qquad \qquad +53\,{t}^{14}+77\,{t}^{15}
+\, \ldots 
\end{eqnarray}
Since  $\, \epsilon \, = 1/5 $ is a non-generic value
(it is of the form $1/m$),
the previous expansion (\ref{expanrealArnold17})
can be compared with the ones corresponding to 
values very close to $1/5$ but not equal, for instance
 $\, \epsilon \, = \, 99/500 $, and $\, \epsilon \, = \, 101/500 $ :
\begin{eqnarray}
\label{expanrealArnold15001}
{\cal A}_{99/500}(t) \, = \, \,t+{t}^{2}+2\,{t}^{3}+3\,{t}^{4}\, +5\,{t}^{5}
+\, 6 \,{t}^{6}+\, 9 \,{t}^{7}+\, 13 \,{t}^{8}+\, 18 \,{t}^{9}
+27\,{t}^{10}+ \,37 \, {t}^{11} + \,62 \, {t}^{12}
+ \,89 \, {t}^{13}\, + \ldots \nonumber
\end{eqnarray}
and :
\begin{eqnarray}
\label{expanrealArnold15002}
{\cal A}_{101/500}(t) \, = \, \,t+{t}^{2}+2\,{t}^{3}+3\,{t}^{4}+5\,{t}^{5}
+6\,{t}^{6}+\, 9 \,{t}^{7}+13\,{t}^{8}+18 \,{t}^{9}
+27\,{t}^{10}+ \,41\,  {t}^{11} + \,66 \, {t}^{12} \,
+ \,85 \, {t}^{13}\, + \ldots \nonumber
\end{eqnarray}
Closer to  $\, \epsilon \, = 1/5 $ one obtains 
for $\, \epsilon \, = \, 999/5000 $, and $\, \epsilon \, = \, 1001/5000 $,
the following expansions :
\begin{eqnarray}
\label{expanrealArnold150011}
{\cal A}_{999/5000}(t) \, = \, \,t+{t}^{2}+2\,{t}^{3}+3\,{t}^{4}\, +5\,{t}^{5}
+\, 6 \,{t}^{6}+\, 9 \,{t}^{7}+\, 13 \,{t}^{8}+\, 18 \,{t}^{9}
+27\,{t}^{10}+ \, 41 \, {t}^{11} + \,66 \, {t}^{12} \, +\,  89 \, t^{13}\,
+  \ldots \nonumber
\end{eqnarray}
and :
\begin{eqnarray}
\label{expanrealArnold150021}
{\cal A}_{1001/5000}(t) \, = \, \,t+{t}^{2}+2\,{t}^{3}+3\,{t}^{4}+5\,{t}^{5}
+6\,{t}^{6}+\, 9 \,{t}^{7}+13\,{t}^{8}+18 \,{t}^{9}
+27\,{t}^{10}+ \,41\,  {t}^{11} + \,66 \, {t}^{12} \, + 89 \, t^{13}\,
+ \, \ldots \nonumber
\end{eqnarray}
Similar expansions, corresponding to
values close to the non-generic value $\, \epsilon \, = \, 1/4$,
are given in Appendix D.
All these  results show that, similarly to the situation for
the customary topological entropy, or the
growth complexity $\, \lambda$ (see Fig.\ref{f:fig1}),
$\, \lambda_{real}\, $ is continuous as a function of $\epsilon$ 
{\em near} the non-generic values of $\, \epsilon \simeq 1/m $, 
however  exactly on these very non-generic values
$\, \lambda_{real}\, $ takes  {\em smaller} values (continuous function up to a 
zero measure set).

\vskip .3cm 

{\bf Remark :} It is natural to compare the expansion 
 corresponding to  $\, \epsilon \,=\, 3/5 $ with the one
 corresponding to  $\, \epsilon \,=\, 1/7 $, since 
 $\, \epsilon \,=1/7 $ and  $\, \epsilon \,=\, 3/5 $ have the {\em same 
topological entropy} (growth complexity $\, \lambda$) associated with 
$1-t-t^2+t^{m+2}$ for $m=7$
 (see relation (\ref{zetamtexte})). One gets  for
 $\, \epsilon \,=\, 3/5 $ :
\begin{eqnarray}
\label{expanrealArnold153sur5}
{\cal A}_{3/5}(t) \, = \, 
t+{t}^{2}+2\,{t}^{3}+3\,{t}^{4}+2\,{t}^{5}
+5\,{t}^{6}+9\,{t}^{7}+8\,{t}^{8}
+11\,{t}^{9}+14\,{t}^{10}+18\,{t}^{11}+24\,{t}^{12} +29\,{t}^{13}\,  
+\,41\,{t}^{14}\, +51\,{t}^{15}\, +
\ldots \nonumber
\end{eqnarray}
and for  $\, \epsilon \,=\, 1/7 $ :
\begin{eqnarray}
\label{expanrealArnold16}
{\cal A}_{1/7}(t) \, = \, 
t\, + t^2\, +2\,{t}^{3}+{t}^{4}+4\,{t}^{5}+7\,{t}^{6}+7\,{t}^{7}
+8\,{t}^{8}+13\,{t}^{9}+16\,{t}^{10}+
22\,{t}^{11} +36\,{t}^{12}+43\,{t}^{13}\,+65\,{t}^{14}\,
+87\,{t}^{15}\, +\, \ldots \nonumber
\end{eqnarray}
 These two expansions {\em do not seem to yield the same 
value for} $\, \lambda_{real}$ ($\lambda_{real}(3/5)\, \simeq \, 51^{1/15} \, \simeq \, 1.2997 $
and $\lambda_{real}(1/7) \, \simeq \, 87^{1/15} \, \simeq \, 1.3468 $)
{\em though they share the same growth complexity}
$\lambda$.

\vskip .3cm 

The expansions of $\, {\cal A}_{\epsilon}(t) \,$
near $\, \epsilon \, = \, 3/5$ are given in Appendix D.

\vskip .3cm 
{\bf Miscellaneous values of $\, \epsilon$.} For 
most of the values of $\, \epsilon\, $ 
the expansions are not long enough to
``guess''  rational expressions (if any ...). 
One can however get some estimates of $\, \lambda_{real}$ that can be compared
with $\, h_{real}$. 

$\bullet\, $ For  $\, \epsilon \, = 21/25 $ one gets the following results:
\begin{eqnarray}
\label{interesting}
{\cal A}_{21/25}(t) \, = \,\,
t+{t}^{2}+2\,{t}^{3}+3\,{t}^{4}+3\,{t}^{5}+6\,{t}^{6}+7\,{t}^{7}+11\,{t}^{8}+12\,{t}^
{9}+21\,{t}^{10}+25\,{t}^{11}+36\,{t}^{12}+45\,{t}^{13}+\, 69\, t^{14}
+\, \ldots  \nonumber
\end{eqnarray}
This expansions  seem to yield the following estimated
value for $\, \lambda_{real}$ ($\lambda_{real}(21/25)\, 
\simeq \, 69^{1/14} \, \simeq \, 1.3531 $
to be compared with (\ref{21sur25})).
In fact, one remarks that this expansion is {\em actually compatible}
 with the expansion
of the rational expression:
\begin{eqnarray}
\label{interesting2}
{\cal A}_{21/25}(t) \, = \,\,
{\frac {t \cdot \left (1+t+{t}^{2}+{t}^{3}-2\,{t}^{4}\right )}{\left (1-t\right )\, (1+t)^{2}
\left (1-t+{t}^{2}-2\,{t}^{3}\right )}}
\end{eqnarray}
One does remark that the rational expression (\ref{interesting2})
has actually the {\em same singularity} that 
 the rational expression  (\ref{compat21sur25}) 
suggested for the real dynamical zeta function $\, \zeta_{21/25}^{real}(t)$.
All the coefficients of the expansion of
 (\ref{interesting2}) are positive (in contrast with  
 (\ref{simpledeuxsurtrois}) given in Appendix C 
 which is ruled out because coefficient $t^{54}$ of its
expansion is negative). If this simple rational expression 
is actually the exact expression for $\, {\cal A}_{21/25}(t) \,$
this would yield the following algebraic value for  $\,
\lambda_{real}$ : 
$\lambda_{real}(21/25)\, = \, h_{real}(21/25) \simeq \, 1.353209964$.
\vskip .2cm
$\bullet\, $ For  $\, \epsilon \, = 3/2 $ one gets the following results:
\begin{eqnarray}
\label{expanrealArnold1610113sur2}
{\cal A}_{3/2}(t) \, = \,\,t+{t}^{2}+2\,{t}^{3}+ \,{t}^{4}
+\,3\, {t}^{5}
+2\,{t}^{6}+3\,{t}^{7}+ 3 \,{t}^{8}
+   \,2\, {t}^{9}+\,3 \, {t}^{10}
 \, +\,3\, {t}^{11} + \,4\, {t}^{12} + \, 3 \, t^{13}
+\, \ldots  \nonumber
\end{eqnarray}

$\bullet\, $ Near  $\, \epsilon \,= 2 \,  $ (for instance for
$ \epsilon \,= 2001/1000 \,  $ or $\, \epsilon \,= 1999/1000 $)
one gets :
\begin{eqnarray}
\label{expanrealArnold16101112001}
{\cal A}_{\epsilon \simeq 2}(t) \, =
\,t+{t}^{2}+2\,{t}^{3}+{t}^{4}+{t}^{5}
+2\,{t}^{6}+{t}^{7}+3\,{t}^{8}+2\,
{t}^{9}+{t}^{10}+3\,{t}^{11}+2\,{t}^{12}+3\,{t}^{13}
+\, \ldots \nonumber
\end{eqnarray}

$\bullet\, $ Some results for  $\, \epsilon \, $ larger than 3
(again obtained with 6000 digits) are given in Appendix D ($ \epsilon
\,= 4 \,  $, 
  $\, \epsilon \,= 5   $,  $\, \epsilon \,= 6   $,
 $\, \epsilon \,= 10 $,  $\, \epsilon \,= 20   $,  $\, \epsilon
\,= 30 \,  $).
These series indicate that an estimated value 
for $\, \lambda_{real}\, $ could correspond to 
 $\, \lambda_{real}\, $ very close to 1 for $\epsilon \, = \, 10$,
and  quite close to 1 for $\epsilon \, = \, 20$.

\subsection{``Real Arnold complexity'' generating functions for 
 $\, \epsilon \, $ large.}
\label{RAClarge}

The examination of Fig.\ref{f:fig12}  shows that 
$\, \lambda_{real} \, $ goes to some non-trivial limit, $\, \lambda_{real} \, \simeq \,
1.429 $, in the 
large  $\epsilon $ limit. Let us examine the expansion
of $\, {\cal A}_{\epsilon}(t)\, $ for 
various increasing values of $\, \epsilon$,
in order to study this $\, \epsilon \rightarrow \, \infty\, $
limit. The expansions of $\, {\cal A}_{\epsilon}(t) \,$
for   $\, \epsilon \,= 40   $,  $\, \epsilon \,= 50  $,
  $\, \epsilon \,= 100 \,  $
are given in Appendix D.

\vskip .3cm 

$ \bullet $ For  $\, \epsilon \,= 500 \,  $ the expansion 
of $\, {\cal A}_{\epsilon}(t) \,$ reads :
\begin{eqnarray}
\label{expanrealArnold1641500}
{\cal A}_{500}(t) \, = \,\, \,\,t+{t}^{2}+2\,{t}^{3}+3\,{t}^{4}
+5\,{t}^{5}\, +\, 8\,{t}^{6}+ \,11\, {t}^{7}
+ \,17 \, {t}^{8}
+ \,24 \, {t}^{9} \, 
+\,  \,35\,  t^{10}+\, \, 47\, t^{11} 
+  \,64\,  t^{12} +  93\, t^{13} +  \,\ldots 
\,  \nonumber
\end{eqnarray}

$ \bullet $  For  $\, \epsilon \,= 1000 \,  $:
\begin{eqnarray}
\label{expanrealArnold1641}
{\cal A}_{1000}(t) \, = \,\, \,\,t+{t}^{2}+2\,{t}^{3}+3\,{t}^{4}
+5\,{t}^{5}\, +8\,{t}^{6}+ \,11\, {t}^{7}
+ \,17 \, {t}^{8}
+ \,24 \, {t}^{9} \, +\,  \,35\,  t^{10}+\, \, 51\, t^{11} 
+  \,72\,  t^{12} +  \, 101\, t^{13}\, + \,\ldots 
\,  \nonumber
\end{eqnarray}

For $\, \epsilon \, $ large the 
expansion of   $\, {\cal A}_{\epsilon}(t) $, the 
generating function 
for the ``real Arnold complexity'',   
is equal, up to order 15,
to (for instance\footnote{These calculations
have to be performed with 
at least 6000 digits. With a number of digits  lower
than 2000 digits
one gets {\em smaller} coefficients: the precision is not large enough to 
distinguish between very close intersection points.}
 for $\, \epsilon \, = \, 20000$) :
\begin{eqnarray}
&&{\cal A}_{20000}(t) \, = \, \, \,
t+{t}^{2}+2\,{t}^{3}+3\,{t}^{4}+5\,{t}^{5}+8\,{t}^{6}+11\,{t}^{7}+17\,{t}^{8}+24\,
{t}^{9}+35\,{t}^{10}+51\,{t}^{11}+72\,{t}^{12}\, 
+105\,{t}^{13}\nonumber \\
&&\qquad \qquad \qquad \qquad \qquad 
+\, 149\,{t}^{14}+214\,{t}^{15}\, + \ldots 
\end{eqnarray}
which actually coincides with the expansion of the 
simple rational expression :
\begin{eqnarray}
\label{expanrealArnold1}
{\cal A}_{\infty}(t) \, \, = \, \,
 \, \, {{ t \cdot (1+t^4)} \over {(1-t^2-t^3-t^5)
\cdot (1-t)}}\, \, = \, \, \,
 {\frac {t \cdot \left (1+{t}^{4}\right )}{ (1-t-{t}^{2})\, +\, {t}^{4} \cdot (1-t+{t}^2 )}}
\end{eqnarray}
This last result has to be compared with 
the equivalent one
for the ``real dynamical zeta function''
$\, \zeta_{\epsilon = \infty}^{real}(t)\, $ (see  (\ref{zetarealinfty})
in section (\ref{seek})).
These two non-trivial rational results, 
for $\epsilon$ large, are in {\em perfect agreement}, yielding the
same {\em
algebraic value} for
the two ``real complexities''  $\,
\, h_{real}\, $ and  $ \,\lambda_{real}\,$, namely  $\,
\, h_{real}\, = \,\,\lambda_{real}\, \simeq \, \, 1.42910832 $.
\vskip .3cm 
All the results displayed in this section seem to 
show that the identification between $\,
\, h_{real}\, $ and  $ \,\lambda_{real}\,$ actually holds.

\vskip .3cm 
{\bf Remark :} Recalling the ``universal'' relation 
(\ref{universal}), or more precisely (\ref{universal2}),
which gives (for $\, \epsilon\, $ generic and for 
$\, \epsilon\,=1/m $,
$\, \epsilon\,=(m+1)/(m+3) $ for $m\, =\, 9\, , \, 13\, , \, 17\, , \,
\ldots$) a ratio $\, \zeta_{\epsilon}(t)/A_{\epsilon}(t) \, $ equal to
$\, (1-t^2)/t $, one can look at the ``real ratio''
$\, \zeta_{\epsilon}^{real}(t)/{\cal A}_{\epsilon}(t)\, $. Of course
for $\epsilon \, <0\, $ this  ``real ratio''
is also equal to $\, (1-t^2)/t $, however in the 
$\epsilon \, \rightarrow \, \infty\, $ limit 
it tends to be equal to $\, (1-t^2)/t/(1+t^4)$.
Therefore one should not expect any simple 
``universal'' relation like (\ref{universal}) 
between $\, \zeta_{\epsilon}^{real}(t)\, $
and $\, {\cal A}_{\epsilon}(t)\, $. 

These various Arnold complexity generating functions  
$\, {\cal A}_{\epsilon}(t)\, $
were associated
with the iteration of the (real or complex) line 
$\, y\, = \, (1-\epsilon)/2 $. One could introduce
a generating function for each line (or fixed curve)
one iterates. The corresponding series become slightly
more difficult to extrapolate but give similar results
in particular the asymptotic values for $\, \lambda_{real}$. 
The sensitivity of the previous analysis, 
according to the chosen curve one iterates,
will be discussed elsewhere.

It would be interesting to see
 if  the ``real dynamical zeta functions'' $\, \zeta_{\epsilon}^{real}(t)\, $, or the
``real degree generating functions'' $\, {\cal A}_{\epsilon}(t)$,
could also be {\em rational} expressions, for  {\em other} 
 values of $\, \epsilon$, or, even, if  these ``real generating
functions'' could be  {\em rational} expressions
for {\em any given} value of $\, \epsilon$. In this last case
there should be an infinite number of  such {\em rational} expressions :
it is clear that 
they could not  all  be ``simple'' like  (\ref{zetarealinfty}) 
or (\ref{expanrealArnold1}).



\section{Conclusion }
\label{Concl}

 The results presented here 
seem to be {\em in agreement with, again, an identification
between} $\, \lambda_{real}$, the (asymptotic)
 ``{\em real Arnold complexity}'', and $\,
h_{real}$, the  (exponential of the) ``{\em real topological
entropy''}. In contrast with the ``universal'' 
behavior of the ``usual'' Arnold complexity,
or topological entropy, displayed in Fig. 1,
 $\, \lambda_{real}\, $ and $h_{real}\, $ are quite involved 
functions of the parameter $\, \epsilon\, $
the birational transformations
 depend on (see Fig. 11 and 
Fig. 12).

We have, however, obtained some remarkable
 {\em rational expressions for
the real dynamical zeta function} $\, \zeta^{real}_{\epsilon}(t)\, $
and for a ``real Arnold complexity''
generating function $\, {\cal A}_{\epsilon}(t)$.
In particular we have obtained  in particular 
two non-trivial {\em rational} expressions
 (\ref{zetarealinfty}) and (\ref{expanrealArnold1}), 
(yielding algebraic values for $\,
\, h_{real}\, $ and  $ \,\lambda_{real}\,$).

There is no simple down-to-earth Markov
partition, symbolic dynamics, or 
hyperbolic systems interpretation of these 
rational results.

\vskip .4cm

\acknowledgments{One of us (JMM) would like to thank 
P. Lochak and J-P. Marco for illuminating
discussions on symbolic dynamics
and on non hyperbolic discrete
 dynamical systems. N. Abarenkova  would like to thank 
the St Petersburg's  administration for a grant.
S. Boukraa would like to thank 
the CMEP for financial support.
}

\pagebreak


\section{Appendix A :  Dynamical zeta functions versus homogeneous
degree generating function for non-generic values}
\label{AppendixA0}

Let us consider here various non-generic values of the form 
$(m-1)/(m+3)$ (with $\, m \, \ge 7$, $m$ odd).

$\bullet$ For $\, \epsilon \, = \, 3/5$ (corresponding to $m\, = \, 7$)
the  homogeneous generating function defined in section (\ref{dynzeta}), reads :
\begin{eqnarray} 
\label{GH3sur5}
G^{Hom}_{3/5}(t)\, \,= \, \, \, \, 
1+2\,t+4\,{t}^{2}+7\,{t}^{3}+12\,{t}^{4}+20\,{t}^{5}+33\,{t}^{6}+54\,{t}^{7}+88\,{t}^{8}+
142\,{t}^{9}+228\,{t}^{10}+366\,{t}^{11}\, + \, \ldots \nonumber 
\end{eqnarray}
which is compatible with the expansion of the rational expression : 
\begin{eqnarray} 
\label{GH3sur5rat}
G^{Hom}_{3/5}(t)\, \,= \, \, \, \, 
{\frac {1-{t}^{10}}{\left (1-t\right ) \cdot \left (1-t-{t}^{2}+{t}^{9}\right )}} \nonumber 
\end{eqnarray}
Recalling a possible  rational expression for the corresponding dynamical zeta
function~\cite{McGuire} :
\begin{eqnarray} 
\label{zet3sur5recall}
\zeta_{3/5}(t)\, \,= \, \, \, \, 
{\frac {1-{t}^{2}}{1-t-{t}^{2}+{t}^{9}}}\, , 
\end{eqnarray}
one immediately verifies that the  ``universal'' relation
(\ref{universal}) actually holds.

$\bullet$ For $\, \epsilon \, = \, 2/3$ (corresponding to $\, m\, = \, 9$)
the  homogeneous generating function defined in section (\ref{dynzeta}), reads :
\begin{eqnarray} 
\label{GH2sur3}
G^{Hom}_{2/3}(t)\, \,= \, \, \, \, 
1+2\,t+4\,{t}^{2}+7\,{t}^{3}+12\,{t}^{4}+20\,{t}^{5}+33\,{t}^{6}+54\,{t}^{7}+88\,{t}^{8}+
143\,{t}^{9}+232\,{t}^{10}+375\,{t}^{11} +\, 605\,{t}^{12}\, + \, \ldots \nonumber 
\end{eqnarray}
This could be the expansion, up to order twelve, of the simple rational expression :
\begin{eqnarray} 
\label{GH2sur3could}
G^{Hom}_{2/3}(t)\, \,= \, \, \, \,
{\frac {1-{t}^{12}}{\left (1-t\right ) \cdot \left (1-t-{t}^{2}+{t}^{11}\right )}} 
\end{eqnarray}
These results  should be compared with the expansion of the dynamical zeta
function. Unfortunately, here, the series for 
the dynamical zeta
function are not sufficiently large to allow any ``safe conjecture''.
A possible exact expression does not seem to be equal to
$\, (1-t^2)/(1-t-{t}^{2}+{t}^{11}) $,
but could be~\cite{McGuire} :
\begin{eqnarray} 
\label{zet2sur3sec}
\zeta_{2/3}(t)\, \,= \, \, \, \, 
{\frac {1-{t}^{2}-{t}^{11}-{t}^{12}-{t}^{13}}{1-t-{t}^{2}+{t}^{11}}}
\qquad \qquad
\hbox{or :} \qquad \qquad {\frac {1-{t}^{2}-{t}^{11}-{t}^{12}}{1-t-{t}^{2}+{t}^{11}}}
\end{eqnarray}
The ``universal'' relation (\ref{universal}) is verified with
(\ref{GH2sur3could}) together with $\, (1-t^2)/(1-t-{t}^{2}+{t}^{11})\, $,
{\em but not  with}
(\ref{GH2sur3could}) together with (\ref{zet2sur3sec}).
One can however imagine that 
the  ``universal'' relation (\ref{universal}) could be slightly
modified 
on some of these $(m-1)/(m+3)$ 
values ($m \, = \, 9\, , \, 13\, , \, \cdots $).
For instance, (\ref{GH2sur3could}) and  (\ref{zet2sur3sec})
verify (up to order twelve) the simple relation :
\begin{eqnarray} 
t \cdot \zeta_{2/3}(t)\, -\, (1-t) \cdot G^{Hom}_{2/3}(t)\, \, +\,  1 \,
-t^{m+2}\, -t^{m+3}\,\,  = \,\,  \, 0 \qquad \qquad \hbox{where :} \qquad m \,
= \, 9
\end{eqnarray}
These calculations need to be revisited.

$\bullet$ For $\, \epsilon \, = \, 5/7$ (corresponding to $m\, = \, 11$)
the  homogeneous generating function defined in section (\ref{dynzeta}), reads :
\begin{eqnarray} 
\label{GH5sur7}
G^{Hom}_{5/7}(t)\, \,= \, \, \, \, 
1+2\,t+4\,{t}^{2}+7\,{t}^{3}+12\,{t}^{4}+20\,{t}^{5}+33\,{t}^{6}+54\,{t}^{7}+88\,{t}^{8}+
143\,{t}^{9}+232\,{t}^{10}\, + 376\,{t}^{11}\, +\,\,609\, t^{12} \,  \ldots \nonumber 
\end{eqnarray}
This could be the expansion of :
\begin{eqnarray} 
\label{GH5sur7could}
G^{Hom}_{5/7}(t)\, \,= \, \, \, \,
{\frac {1-{t}^{14}}{\left (1-t\right ) \cdot \left (1-t-{t}^{2}+{t}^{13}\right )}} \nonumber 
\end{eqnarray}

$\bullet$ For $\, \epsilon \, = \, 3/4$ (corresponding to $m\, = \, 13$)
the  homogeneous generating function reads :
\begin{eqnarray} 
\label{GH3sur4}
G^{Hom}_{3/4}(t)\, \,= \, \, \, \, 
1+2\,t+4\,{t}^{2}+7\,{t}^{3}+12\,{t}^{4}+20\,{t}^{5}
+33\,{t}^{6}+54\,{t}^{7}+88\,{t}^{8}+
143\,{t}^{9}+232\,{t}^{10}+376\,{t}^{11}\,+609\, t^{12} \, + \, \ldots \nonumber 
\end{eqnarray}
This series is not large enough. It could be the expansion of the simple expression :
\begin{eqnarray} 
\label{GH3sur4could}
G^{Hom}_{3/4}(t)\, \,= \, \, \, \,
{\frac {1-{t}^{16}}{\left (1-t\right )
 \cdot \left (1-t-{t}^{2}+{t}^{15}\right )}} \nonumber 
\end{eqnarray}

\pagebreak

\section{Appendix B : Number of real fixed points
 of the $\, P$-type,  $\, Q$-type
and  $\, R$-type.}
\label{AppendixA}

Let us just give the number  of  {\em real} $n$-th
 cycles of the $P$-type, $Q$-type and
$R$-type for miscellaneous values of $\epsilon$ in increasing order. 

For $\epsilon < 0$ (and $\epsilon \ne -1$) one gets : 
\begin{table}[t]
\begin{tabular}{||c|c|c|c|c|c|c|c|c|c|c|c|c|c|c|c|c|c|c||} \hline
 $n$      & 1 & 2 & 3 & 4 & 5 & 6 & 7 & 8 & 9 & 10& 11 & 12 & 13 & 14 & 15& 16& 17& 18\\ \hline
 $P_n$    & 1 & 0 & 1 & 1 & 2 & 2 & 4 & 4 & 6 & 8 & 12 & 12 & 20 & 24 & 30 & 38& 54& 65 \\ \hline
 $Q_n$    & 0 & 0 & 0 & 0 & 0 & 0 & 0 & 1 & 0 & 1 & 0  & 3  & 0  & 4 & 0  & 9 & 0 & 13 \\ \hline
 $R_n$    & 0 & 0 & 0 & 0 & 0 & 0 & 0 & 0 & 2 & 2 & 6  & 10 & 20 & 30
 & 60 & 88 & 156 & 238 \\ \hline
 $T_n$    & 1 & 0 & 1 & 1 & 2 & 2 & 4 & 5 & 8 & 11& 18 & 25 & 40 & 58
 & 90  & 135 & 210 & 316 \\ \hline
\end{tabular}
\caption{Number  of  {\em real} $n$-th cycles of the $P$-type, $Q$-type and
$R$-type for $\epsilon < 0$.
\label{latable2}
}
\end{table}

For $\, \epsilon = 11/100,\,  1/4,\,  52/100,\,  9/10,\,  11/10,\,  5,\,  10,\,  50,\,  100 $,
 and  $\, 20000$,
one gets the following tables :
\begin{table}
\begin{tabular}{||c|c|c|c|c|c|c|c|c|c|c|c|c|c|c|c|c|c|c||} \hline
 $n$      & 1 & 2 & 3 & 4 & 5 & 6 & 7 & 8 & 9 & 10& 11& 12& 13& 14& 15& 16& 17& 18\\ \hline
 $P_n$    & 1 & 0 & 1 & 1 & 2 & 1 & 2 & 1 & 4 & 5 & 8 & 5 & 10& 11& 14& 14& 20& 21 \\ \hline
 $Q_n$    & 0 & 0 & 0 & 0 & 0 & 0 & 0 & 0 & 0 & 1 & 0 & 3 & 0 & 3 & 0  & 5 & 0& 5 \\ \hline
 $R_n$    & 0 & 0 & 0 & 0 & 0 & 0 & 0 & 0 & 0 & 0 & * & * & * & * & * & * & *  & * \\ \hline
 $T_n$    & 1 & 0 & 1 & 1 & 2 & 1 & 2 & 1 & 4 & 6 & * & * & * & * & * & *& * & * \\ \hline
\end{tabular}
\caption{Number  of  {\em real} $n$-th cycles of the $P$-type, $Q$-type and
$R$-type for $\epsilon = 11/100$.
\label{latable4}
}
\end{table}
\begin{table}
\begin{tabular}{||c|c|c|c|c|c|c|c|c|c|c|c|c|c||} \hline
 $n$      & 1 & 2 & 3 & 4 & 5 & 6 & 7 & 8 & 9 & 10& 11& 12& 13\\ \hline
 $P_n$    & 1 & 0 & 1 & 1 & 2 & 0 & 1 & 0 & 3 & 1 & 4 & 2 & 8  \\ \hline
 $Q_n$    & 0 & 0 & 0 & 0 & 0 & 0 & 0 & 1 & 0 & 1 & 0 & 2 & 0  \\ \hline
 $R_n$    & 0 & 0 & 0 & 0 & 0 & 0 & 0 & 0 & 0 & 0 & 0 & 0 & 0  \\ \hline
 $T_n$    & 1 & 0 & 1 & 1 & 2 & 0 & 1 & 1 & 3 & 2 & 4 & 4 & 8 \\ \hline
\end{tabular}
\caption{Number  of  {\em real} $n$-th cycles of the $P$-type, $Q$-type and
$R$-type for $\epsilon \, = \, 1/4$.
\label{latableunsurquatre}
}
\end{table}
\begin{table}
\begin{tabular}{||c|c|c|c|c|c|c|c|c|c|c|c|c|c|c|c||} \hline
 $n$      & 1 & 2 & 3 & 4 & 5 & 6 & 7 & 8 & 9 & 10& 11& 12& 13& 14& 15\\ \hline
 $P_n$    & 1 & 0 & 1 & 1 & 2 & 0 & 2 & 0 & 4 & 1 & 4 & 2 & 6 & 2 & 8 \\ \hline
 $Q_n$    & 0 & 0 & 0 & 0 & 0 & 0 & 0 & 1 & 0 & 1 & 0 & 2 & 0 & 4 & 0 \\ \hline
 $R_n$    & 0 & 0 & 0 & 0 & 0 & 0 & 0 & 0 & 0 & 0 & 2 & 8 & 10 & * & * \\ \hline
 $T_n$    & 1 & 0 & 1 & 1 & 2 & 0 & 2 & 1 & 4 & 2 & 6 & 12 & 16 & * & *\\ \hline
\end{tabular}
\caption{Number  of  {\em real} $n$-th cycles of the $P$-type, $Q$-type and
$R$-type for $\epsilon \, = \, 52/100$.
\label{latable}
}
\end{table}
\begin{table}
\begin{tabular}{||c|c|c|c|c|c|c|c|c|c|c|c|c|c|c|c|c|c|c||} \hline
 $n$      & 1 & 2 & 3 & 4 & 5 & 6 & 7 & 8 & 9 & 10& 11& 12& 13& 14& 15& 16& 17& 18\\ \hline
 $P_n$    & 1 & 0 & 1 & 1 & 0 & 0 & 2 & 0 & 0 & 1 & 4 & 0 & 2 & 1 & 6& 1 & 6  & 3 \\ \hline
 $Q_n$    & 0 & 0 & 0 & 0 & 0 & 0 & 0 & 1 & 0 & 1 & 0 & 2 & 0 & 2 & 0
 & 4 & 0 & 5 \\ \hline
 $R_n$    & 0 & 0 & 0 & 0 & 0 & 0 & 0 & 0 & 0 & 0 & * & * & * & * & * & * & *  & *   \\ \hline
 $T_n$    & 1 & 0 & 1 & 1 & 0 & 0 & 2 & 1 & 0 & 2 & * & * & * & * & * & *& *  & *  \\ \hline
\end{tabular}
\caption{Number  of  {\em real} $n$-th cycles of the $P$-type, $Q$-type and
$R$-type for $\epsilon = 9/10$.
\label{latable49}
}
\end{table}
\begin{table}
\begin{tabular}{||c|c|c|c|c|c|c|c|c|c|c|c|c|c|c|c|c|c||} \hline
 $n$      & 1 & 2 & 3 & 4 & 5 & 6 & 7 & 8 & 9 & 10& 11& 12& 13& 14& 15& 16& 17\\ \hline
 $P_n$    & 1 & 0 & 1 & 0 & 0 & 0 & 2 & 0 & 0 & 1 & 2 & 0 & 2 & 1 & 2 & 1 & 4  \\ \hline
 $Q_n$    & 0 & 0 & 0 & 0 & 0 & 0 & 0 & 0 & 0 & 1 & 0 & 0 & 0 & 0 & 0 & 1 & 0 \\ \hline
 $R_n$    & 0 & 0 & 0 & 0 & 0 & 0 & 0 & 0 & 0 & 0 & 0 & * & * & * & * & * & *  \\ \hline
 $T_n$    & 1 & 0 & 1 & 0 & 0 & 0 & 2 & 0 & 0 & 2 & 2 & * & * & * & * & * & * \\ \hline
\end{tabular}
\caption{Number  of  {\em real} $n$-th cycles of the $P$-type, $Q$-type and
$R$-type for $\epsilon \, = \, 11/10$.
\label{latable311}
}
\end{table}
\begin{table}
\begin{tabular}{||c|c|c|c|c|c|c|c|c|c|c|c|c|c|c|c|c|c|c||} \hline
 $n$      & 1 & 2 & 3 & 4 & 5 & 6 & 7 & 8 & 9 & 10& 11& 12& 13& 14& 15& 16& 17& 18\\ \hline
 $P_n$    & 1 & 0 & 1 & 0 & 0 & 0 & 0 & 0 & 0 & 0 & 2 & 0 & * & 1 & 0 & 0 & 2 & 0 \\ \hline
 $Q_n$    & 0 & 0 & 0 & 0 & 0 & 0 & 0 & 0 & 0 & 0 & 0 & 0 & 0 & 1 & 0 & 0 & 0 & 0\\ \hline
 $R_n$    & 0 & 0 & 0 & 0 & 0 & 0 & 0 & 0 & 0 & 0 & 0 & * & * & * & * & * & * & *\\ \hline
 $T_n$    & 1 & 0 & 1 & 0 & 0 & 0 & 0 & 0 & 0 & 0 & 2 & * & * & * & * & * & * & *\\ \hline
\end{tabular}
\caption{Number  of  {\em real} $n$-th cycles of the $P$-type, $Q$-type and
$R$-type for $\epsilon \, = \, 5$.
\label{latable31}
}
\end{table}
\begin{table}
\begin{tabular}{||c|c|c|c|c|c|c|c|c|c|c|c|c|c|c|c|c|c|c||} \hline
 $n$      & 1 & 2 & 3 & 4 & 5 & 6 & 7 & 8 & 9 & 10& 11& 12& 13& 14& 15& 16& 17& 18\\ \hline
 $P_n$    & 1 & 0 & 1 & 0 & 2 & 0 & 0 & 1 & 0 & 0 & 2 & 0 & 2 & 1 & 0 & 0 & 2 & 1 \\ \hline
 $Q_n$    & 0 & 0 & 0 & 0 & 0 & 0 & 0 & 1 & 0 & 0 & 0 & 0 & 0 & 1 & 0 & 0 & 0 & 1\\ \hline
 $R_n$    & 0 & 0 & 0 & 0 & 0 & 0 & 0 & 0 & 0 & 0 & 0 & * & * & * & * & * & *  & *\\ \hline
 $T_n$    & 1 & 0 & 1 & 0 & 2 & 0 & 0 & 2 & 0 & 0 & 2 & * & * & * & * & * & * & *\\ \hline
\end{tabular}
\caption{Number  of  {\em real} $n$-th cycles of the $P$-type, $Q$-type and
$R$-type for $\epsilon \, = \, 10$.
\label{latable310}
}
\end{table}
\begin{table}
\begin{tabular}{||c|c|c|c|c|c|c|c|c|c|c|c|c|c|c|c|c|c|c||} \hline
 $n$      & 1 & 2 & 3 & 4 & 5 & 6 & 7 & 8 & 9 & 10& 11& 12& 13& 14& 15& 16& 17& 18\\ \hline
 $P_n$    & 1 & 0 & 1 & 0 & 2 & 0 & 2 & 1 & 2 & 0 & 4 & 1 & 4 & 1 & 2 & 2 & 6 & 3 \\ \hline
 $Q_n$    & 0 & 0 & 0 & 0 & 0 & 0 & 0 & 1 & 0 & 1 & 0 & 1 & 0 & 2 & 0 & 3 & 0 & 3\\ \hline
 $R_n$    & 0 & 0 & 0 & 0 & 0 & 0 & 0 & 0 & 0 & 2 & 0 & * & * & * & * & * & *  & *\\ \hline
 $T_n$    & 1 & 0 & 1 & 0 & 2 & 0 & 2 & 2 & 2 & 3 & 4 & * & * & * & * & * & * & *\\ \hline
\end{tabular}
\caption{Number  of  {\em real} $n$-th cycles of the $P$-type, $Q$-type and
$R$-type for $\epsilon \, = \, 50$.
\label{latable32}
}
\end{table}
\begin{table}
\begin{tabular}{||c|c|c|c|c|c|c|c|c|c|c|c|c|c|c|c|c|c|c||} \hline
 $n$      & 1 & 2 & 3 & 4 & 5 & 6 & 7 & 8 & 9 & 10& 11& 12& 13& 14& 15& 16& 17& 18\\ \hline
 $P_n$    & 1 & 0 & 1 & 0 & 2 & 0 & 2 & 1 & 2 & 0 & 4 & 1 & 6 & 1 & 6 & 2 & 8 & 3 \\ \hline
 $Q_n$    & 0 & 0 & 0 & 0 & 0 & 0 & 0 & 1 & 0 & 1 & 0 & 3 & 0 & 2 & 0 & 3 & 0 & 6\\ \hline
 $R_n$    & 0 & 0 & 0 & 0 & 0 & 0 & 0 & 0 & 0 & 2 & 0 & * & * & * & * & * & *  & *\\ \hline
 $T_n$    & 1 & 0 & 1 & 0 & 2 & 0 & 2 & 2 & 2 & 3 & 4 & * & * & * & * & * & * & *\\ \hline
\end{tabular}
\caption{Number  of  {\em real} $n$-th cycles of the $P$-type, $Q$-type and
$R$-type for $\epsilon \, = \, 100$.
\label{latable33}
}
\end{table}
\begin{table}
\begin{tabular}{||c|c|c|c|c|c|c|c|c|c|c|c|c|c|c|c|c||} \hline
 $n$      & 1 & 2 & 3 & 4 & 5 & 6 & 7 & 8 & 9 & 10& 11& 12& 13& 14& 15& 16\\ \hline
 $P_n$    & 1 & 0 & 1 & 0 & 2 & 0 & 2 & 1 & 2 & 0 & 4 & 1 & 6 & 1 & 6 & 2 \\ \hline
 $Q_n$    & 0 & 0 & 0 & 0 & 0 & 0 & 0 & 1 & 0 & 1 & 0 & 3 & 0 & 4 & 0 & 7\\ \hline
 $R_n$    & 0 & 0 & 0 & 0 & 0 & 0 & 0 & 0 & 0 & 2 & 0 & 2 & * & * & * & * \\ \hline
 $T_n$    & 1 & 0 & 1 & 0 & 2 & 0 & 2 & 2 & 2 & 3 & 4 & 6 & $\ge 8$ & $\ge 9$ & $\ge 14$ & $\ge 17$ \\ \hline
\end{tabular}
\caption{Number  of  {\em real} $n$-th cycles of the $P$-type, $Q$-type and
$R$-type for $\epsilon \, = \, 20000$.
\label{latable3}
}
\end{table}

\section{Appendix C : expansions of some
 real dynamical zeta functions}
\label{AppendixB1}

 Let us just give  some additional
 expansions for $\, \zeta^{real}_{\epsilon}(t) \,$
for  increasing values of $\, \epsilon$.

For  $\epsilon=9/50$,
one obtains the following
expansions for $\, \zeta^{real}_{\epsilon}(t) \,$ :
\begin{eqnarray}
\label{9sur50}
&&\zeta^{real}_{9/50}(t) \,\, = \,\, \,
{\frac {1}{\left (1-t\right )
\left (1-{t}^{3}\right )\left (1-{t}^{4}\right )\left (1
-{t}^{5}\right )^{2}\left (1-{t}^{6}\right )
\left (1-{t}^{7}\right )^{2}\left (1-{t}^
{8}\right )^{3}\left (1-{t}^{9}\right )^{4}
\left (1-{t}^{10}\right )^{3}\left (1-{t}^
{11}\right )^{8}}}
\, \cdots \nonumber \\
&& \qquad \qquad \,\, = \,\, \,
1+t+{t}^{2}+2\,{t}^{3}+3\,{t}^{4}+5\,{t}^{5}
+7\,{t}^{6}+10\,{t}^{7}+16\,{t}^{8}+24\,
{t}^{9}+34\,{t}^{10}+52\,{t}^{11}
+ \, \ldots \nonumber
\end{eqnarray}
 yielding the following ``rough'' approximation
 for $\, h_{real} \,$ :
 $\, h_{real} \, \simeq \,  (52)^{1/11} \, \simeq \, 1.432 $.
For  the ``non-generic'' value  $\epsilon=1/5$, $\,
 \zeta^{real}_{1/5}(t) \,$
reads  :
\begin{eqnarray}
\label{unsur5}
&&\zeta^{real}_{1/5}(t) \,\, = \,\, \,
\, {\frac {1}{\left (1-t\right )
\left (1-{t}^{3}\right )
\left (1-{t}^{4}\right )\left (1
-{t}^{5}\right )^{2}\left (1-{t}^{6}\right )
\left (1-{t}^{7}\right ) \left (1-{t}^{9}\right )^{4}
(1-{t}^{10} )^{2} (1-{t}^{11} )^{5} \, (1-t^{12} )^{4}}} \, \cdots \nonumber \\
&& \qquad \qquad\,\, = \,\, \,
1+t+{t}^{2}+2\,{t}^{3}+3\,{t}^{4}+5\,{t}^{5}
+7\,{t}^{6}+9\,{t}^{7}+12\,{t}^{8}+20\,{t}^{9}
+28\,{t}^{10}+39\,{t}^{11}+55\,{t}^{12}
+ \, \ldots \nonumber
\end{eqnarray}
 yielding the following ``rough'' approximation
 for $\, h_{real} \,$ :
 $\, h_{real} \, \simeq \,  (55)^{1/12} \, \simeq \,1.3964 $.
For  $\epsilon=1/5$ the previous $\, Q_n$'s and $\, R_n$'s are equal to 
zero up to order twelve. The exponents in (\ref{unsur5}) are thus the $\, P_n$'s.

For    $\, \epsilon\, =\, 31/125$,  
$\, \epsilon \, = \, 12/25$, $\epsilon=66/125$,   $\epsilon=2/3$,
$\epsilon=17/25$,  
 $\epsilon=3/4$, 
 $\epsilon=3/2$, 
one obtains, respectively, the following
expansions for $\, \zeta^{real}_{\epsilon}(t) \,$ :
\begin{eqnarray}
\label{31sur125}
&&\zeta^{real}_{31/125}(t) \,\, = \,\, \,
\, {\frac {1}{\left (1-t\right )
\left (1-{t}^{3}\right )\left (1-{t}^{4}\right )\left (1
-{t}^{5}\right )^{2}\left (1-{t}^{6}\right )
\left (1-{t}^{7}\right )^{2}\left (1-{t}^
{8}\right )\left (1-{t}^{9}\right )^{4}
\left (1-{t}^{10}\right )^{5}\left (1-{t}^{11}
\right )^{12}}}\, \cdots \nonumber \\
&& \qquad \qquad\,\, = \,\, \,
1+t+{t}^{2}+2\,{t}^{3}+3\,{t}^{4}+5\,{t}^{5}
+7\,{t}^{6}+10\,{t}^{7}+14\,{t}^{8}+22\,
{t}^{9}+34\,{t}^{10}+54\,{t}^{11}
+ \, \ldots \nonumber
\end{eqnarray}
 yielding the  approximation
 for $\, h_{real} \,$ :
 $\, h_{real} \, \simeq \,  (54)^{1/11} \, \simeq \,1.437 $,
\begin{eqnarray}
\label{12sur25}
&&\zeta^{real}_{12/25}(t) \,\, = \,\, \,
{\frac {1}{\left (1-t\right )\left (1-{t}^{3}\right )\left (1-{t}^{4}
\right ) \left (1-{t}^{5}\right )^2 \left
(1-{t}^{7}\right )^2 \left (1-{
t}^{8}\right )\left (1-{t}^{9}\right )^{6} \left (1-{t}^{10}\right )^{5}
\left (1-{t}^{11}\right )^{10} }} \, \cdots
\, \nonumber \\
&& \qquad \qquad \,\, = \,\, \,1+t+{t}^{2}
+2\,{t}^{3}+3\,{t}^{4}+5\,{t}^{5}\,
+6\,{t}^{6}+9\,{t}^{7}+13\,{t}^{8}+22\,{t
}^{9}+33\,{t}^{10}+49\,{t}^{11}
+ \, \ldots \nonumber
\end{eqnarray}
for $\, \epsilon \, = \, 12/25$, yielding the following rough approximation
 for $\, h_{real} \,$ :
 $\, h_{real} \, \simeq \,  (49)^{1/11} \, \simeq \, 1.424 $,
\begin{eqnarray}
\label{66sur125}
&&\zeta^{real}_{66/125}(t) \,\, = \,\, \,
{\frac {1}{\left (1-t\right )\left (1-{t}^{3}\right )
\left (1-{t}^{4}\right )\left (1
-{t}^{7}\right )^{2}\left (1-{t}^{8}\right )
\left (1-{t}^{9}\right )^{4}\left (1-{t}^
{10}\right )^{2}\left (1-{t}^{11}\right )^{4}}} \, \cdots
\, \nonumber \\
&& \qquad \qquad\,\, = \,\, \,
1+t+{t}^{2}+2\,{t}^{3}+3\,{t}^{4}+3\,{t}^{5}
+4\,{t}^{6}+7\,{t}^{7}+9\,{t}^{8}+14\,{t}
^{9}+19\,{t}^{10}+27\,{t}^{11}
+ \, \ldots \nonumber
\end{eqnarray}
for $\, \epsilon \, = \, 66/125$, yielding 
 $\, h_{real} \, \simeq \,  (27)^{1/11} \, \simeq \, 1.349  $.

For $\, \epsilon \, = \, 2/3$ (that is $(m-1)/(m+3)\, $ for $\, m\, = \, 9$)
the real dynamical zeta function reads :
\begin{eqnarray}
\label{deuxsurtrois}
&&\zeta^{real}_{2/3}(t) \,\, = \,\, \,
{\frac {1}{ (1-t) (1-t^3)
\left (1-{t}^{4}\right )\left (1
-{t}^{7}\right )^{2}\left (1-{t}^{8}\right )
\left (1-{t}^{9}\right )^{2}\left (1-{t}^
{10}\right )^{2}\left (1-{t}^{11}\right )^{4}\, (1-{t}^{12})^{2} }} \, \cdots
\,  \\
&& \qquad \qquad\,\, = \,\, \,
1+t+{t}^{2}+2\,{t}^{3}+3\,{t}^{4}+3\,{t}^{5}+4\,{t}^{6}+7\,{t}^{7}+9\,{t}^{8}+12\,{t}
^{9}+17\,{t}^{10}+25\,{t}^{11}+32\,{t}^{12}
+ \, \ldots \nonumber
\end{eqnarray}
 yielding  $\, h_{real} \, \simeq \,  (32)^{1/12} \, \simeq \, 1.3348  $.
Let us note that one must be
careful converting systematically 
a  series to a rational function (Pade approximation).
Up to order twelve, expansion (\ref{deuxsurtrois}) is in agreement with the
expansion of the following simple rational expression :
\begin{eqnarray}
\label{simpledeuxsurtrois}
{\frac {1+t+{t}^{3}-{t}^{6}}{1-{t}^{2}-2\,{t}^{4}+{t}^{5}-{t}^{6}}} 
\,\, = \,\, \,
{\frac {\left (1+t+{t}^{3}-{t}^{6}\right ) \cdot \left (1-t\right
)}{1-t-{t}^{2}\, +{t}^{3}\cdot \left (
1-t+t^2\right )^{2}}}
\end{eqnarray}
which is reminiscent of the exact expression (\ref{zetarealinfty}). 
However, one easily finds that the coefficients of $\, t^{54}\, $ 
in (\ref{simpledeuxsurtrois})
becomes negative 
(the coefficients grow like $\, \simeq \, (-1.5252)^N$). 
Expression (\ref{simpledeuxsurtrois}) {\em cannot} be 
the exact expression of a (real) dynamical zeta function.

For $\, \epsilon \, = \, 17/25$, the real dynamical zeta function
reads :
\begin{eqnarray}
\label{17sur25}
&&\zeta^{real}_{17/25}(t) \,\, = \,\, \,
{\frac {1}{\left (1-t\right )\left (1-{t}^{3}\right )
\left (1-{t}^{4}\right )\left (1
-{t}^{7}\right )^{2}\left (1-{t}^{8}\right )
\left (1-{t}^{9}\right )^{2}\left (1-{t}^
{10}\right )^{2}\left (1-{t}^{11}\right )^{4}}} \,\, \cdots 
\, \nonumber \\
&& \qquad \qquad \,\, = \,\, \,
1+t+{t}^{2}+2\,{t}^{3}+3\,{t}^{4}+3\,{t}^{5}
+4\,{t}^{6}+7\,{t}^{7}+9\,{t}^{8}+12\,{t}
^{9}+17\,{t}^{10}+25\,{t}^{11}
+ \, \ldots \nonumber
\end{eqnarray}
 yielding : $\, h_{real} \, \simeq \,  (25)^{1/11} \, \simeq \,1.3399  $,

For the non-generic value 
$\, \epsilon \, = \, 3/4$ (that is $(m-1)/(m+3)\, $ for $\, m\, = \, 13$)
the real dynamical zeta function reads :
\begin{eqnarray}
\label{troissurquatre}
&&\zeta^{real}_{3/4}(t) \,\, = \,\, \,
{\frac {1}{ (1-t) (1-t^3)
\left (1-{t}^{4}\right )\left (1
-{t}^{7}\right )^{2}\left (1-{t}^{8}\right )
\left (1-{t}^{9}\right )^{3}\left (1-{t}^
{10}\right ) (1-t^{11} )^{2}\, (1-t^{12}) }} \, \cdots
\, \nonumber \\
&& \qquad \qquad\,\, = \,\, \,
1+t+{t}^{2}+2\,{t}^{3}+3\,{t}^{4}+3\,{t}^{5}+4\,{t}^{6}+7\,{t}^{7}+9\,{t}^{8}+13\,{t}
^{9}+17\,{t}^{10}+23\,{t}^{11}+30\,{t}^{12}
+ \, \ldots \nonumber
\end{eqnarray}
 yielding 
 $\, h_{real} \, \simeq \,  (30)^{1/12} \, \simeq \, 1.3277  $.

Finally, for  $\, \epsilon \, = \, 3/2$, the real dynamical zeta function
reads :
\begin{eqnarray}
\label{3sur2}
&&\zeta^{real}_{3/2}(t) \,\, = \,\, \,
{\frac {1}{\left (1-t\right )\left (1-{t}^{3}\right )
\left (1-{t}^{7}\right )^{2}
\left (1-{t}^{10}\right )^{2}}}
\, \,\cdots \nonumber \\
&& \qquad \qquad \,\, = \,\, \,
1+t+{t}^{2}+2\,{t}^{3}+2\,{t}^{4}+2\,{t}^{5}
+3\,{t}^{6}+5\,{t}^{7}+5\,{t}^{8}+6\,{t}
^{9}+10\,{t}^{10}+10\,{t}^{11}
+ \, \ldots \nonumber
\end{eqnarray}
 yielding  :
 $\, h_{real} \, \simeq \,  (10)^{1/11} \, \simeq \,1.233 $.

\vskip 2cm 

\section{Appendix D : Expansions of the
 ``real Arnold complexity'' generating functions}
\label{AppendixB}

Let us give here a few  expansions for the ``real Arnold complexity''
generating functions $\, {\cal A}_{\epsilon}(t)$. Let us first give
the expansion of $\, {\cal A}_{\epsilon}(t)\, $
 corresponding to 
$\epsilon \, = \, 2/3\, $ in order to compare it with 
 (\ref{deuxsurtrois}) and   (\ref{simpledeuxsurtrois}) :
\begin{eqnarray}
\label{expanrealArnoldsimpledeuxsurtrois}
{\cal A}_{2/3}(t) \, = \, \,
t+{t}^{2}+2\,{t}^{3}+3\,{t}^{4}+3\,{t}^{5}+6\,{t}^{6}+7\,{t}^{7}
+11\,{t}^{8}+14\,{t}^{9}+21\,{t}^{10}+29\,{t}^{11}+37\,{t}^{12}+51\,{t}^{13}
\, +\,  \ldots \nonumber
\end{eqnarray}
yielding the following estimation for $\, \lambda_{real} \, \simeq\,
(51)^{1/13}\, \simeq \,1.3531\,  $ to be compared with  $\, h_{real} \, \simeq\,
(32)^{1/12}\, \simeq \, 1.3348 $   from  (\ref{deuxsurtrois}).
The coefficients of the
 expansions of   $\, \zeta_{2/3}(t)$, and $\, {\cal A}_{2/3}(t) $,
are very close. Up to order ten, the ratio  
$\, \zeta_{2/3}(t)/{\cal A}_{2/3}(t)\, $
coincides with the expansion of :
\begin{eqnarray}
 {{\zeta_{2/3}(t)} \over {{\cal A}_{2/3}(t)}}\,\, \simeq \, \,\,
{{ 1-t^2} \over { t }} \cdot {\frac {1-{t}^{5}}{1-2\,{t}^{4}+{t}^{5}}}
\end{eqnarray}
The expansion of $\, {\cal A}_{\epsilon}(t)\, $
 corresponding to the non-generic value
$\, \epsilon \, = \, 3/4\, $ (that is $(m-1)/(m+3)$ for $m\, = \, 13$) reads:
\begin{eqnarray}
\label{expanrealArnoldsimpletroissurquatre}
{\cal A}_{3/4}(t) \, = \, \,
1+t+{t}^{2}+2\,{t}^{3}+3\,{t}^{4}+3\,{t}^{5}+6\,{t}^{6}
+7\,{t}^{7}+11\,{t}^{8}+12\,{t}^{9}+21\,{t}^{10}\, +27\cdot t^{11}
\, + \, 36\, {t}^{12}\, +\, 47\,{t}^{13} +\,  \ldots 
\end{eqnarray}
The  expansions of   $\, \zeta_{3/4}(t)$ and $\, {\cal A}_{3/4}(t)\, $
are again very close.
Expansion (\ref{expanrealArnoldsimpletroissurquatre})
yields the following estimation for $\, \lambda_{real} \, \simeq\,
(47)^{1/13}\, \simeq \, 1.3446\,  $ to be compared with  $\, h_{real} \, \simeq\,
(30)^{1/12}\, \simeq \, 1.3276 $   from  (\ref{troissurquatre}).

Let us now give the expansion of $\, {\cal A}_{\epsilon}(t)\, $
 corresponding to 
values very close to the non-generic value $\, 1/4$, for instance
$\epsilon \, = \, 99/400\, $ and $\epsilon \, = \, 101/400\,  $ :
\begin{eqnarray}
\label{expanrealArnold14001}
{\cal A}_{99/400}(t) \, = \, \,t+{t}^{2}+2\,{t}^{3}+3\,{t}^{4}+5\,{t}^{5}
+6 \,{t}^{6}+\, 9 \,{t}^{7}+13\,{t}^{8}+22\,{t}^{9}
+33\,{t}^{10}+ \,47\,  {t}^{11} + \,70\, {t}^{12} \, + \, 101 \,
t^{13}\, +\,  \ldots \nonumber
\end{eqnarray}
and :
\begin{eqnarray}
\label{expanrealArnold14002}
{\cal A}_{101/400}(t) \, = \, \,t+{t}^{2}+2\,{t}^{3}+3\,{t}^{4}+5\,{t}^{5}
+6 \,{t}^{6}+\, 9 \,{t}^{7}+13\,{t}^{8}+22\,{t}^{9}
+33\,{t}^{10}+ \,47\,  {t}^{11} + \,70\, {t}^{12} \, + 109 \, t^{13}
\, +\, \ldots \nonumber
\end{eqnarray}

Near the non-generic  value
 $\, \epsilon \, = \, 3/5$, for instance for 
$\, \epsilon \, = \, 299/500$ and
 $\, \epsilon \, = \, 301/500$, one gets :
\begin{eqnarray}
\label{expanrealArnold15299}
{\cal A}_{299/500}(t) \, = \, \, 
t+{t}^{2}+2\,{t}^{3}+3\,{t}^{4}+3\,{t}^{5}+6\,{t}^{6}
+11\,{t}^{7}+\,11 \, {t}^{8}
+\,16\, {t}^{9}+ \,21 \, {t}^{10}
+ \,29 \, {t}^{11}+ \,42 \, {t}^{12} \, + \,57 \,{t}^{13}\, +\, 
\ldots \nonumber
\end{eqnarray}
and :
\begin{eqnarray}
\label{expanrealArnold15}
{\cal A}_{301/500}(t) \, = \, \, 
t+{t}^{2}+2\,{t}^{3}+3\,{t}^{4}+3\,{t}^{5}+6\,{t}^{6}
+11\,{t}^{7}+\,11 {t}^{8}
+\,16 {t}^{9}+ \,21 {t}^{10}+ \,29\, {t}^{11}
+ \,42\, {t}^{12} \,  + \,57 \,{t}^{13}\,+
\ldots \nonumber
\end{eqnarray}

For the non-generic value $\, \epsilon \, = 1/10 $ one obtains the following results :
\begin{eqnarray}
\label{expanrealArnold16101}
{\cal A}_{1/10}(t) \, = \, \,t+{t}^{2}+2\,{t}^{3}+ \,{t}^{4}+\,3\, {t}^{5}
+8\,{t}^{6}+9\,{t}^{7}+ 11 \,{t}^{8}
+  16 \,{t}^{9}+\, \,21 \, {t}^{10} \, 
+\,31 \, {t}^{11} +\,48\, {t}^{12} + \,58\,  t^{13}
+\, \ldots \nonumber
\end{eqnarray}

$\bullet$ {\bf $\epsilon >3$.} Let us finally give
 some results for  $\, \epsilon \, $ larger than 3
(again obtained with 6000 digits).

For  $\, \epsilon \,= 4 \,  $:
\begin{eqnarray}
\label{expanrealArnold161011}
{\cal A}_{4}(t) \, = \, \,t+{t}^{2}+2\,{t}^{3}+\,{t}^{4}+\,\, {t}^{5}
+\, 2 \,{t}^{6}+\, \,{t}^{7}+ \,{t}^{8}
+ 2 \,{t}^{9}+3\,{t}^{10}  \, +\,{t}^{11} +\, 2\,{t}^{12} +3 \, t^{13}\, + \, t^{14}
+\, \ldots  \nonumber 
\end{eqnarray}
For  $\, \epsilon \,= 5 \,  $:
\begin{eqnarray}
\label{expanrealArnold1610111}
{\cal A}_{5}(t) \, = \, \,t+{t}^{2}+2\,{t}^{3}+\,{t}^{4}+\,\, {t}^{5}
+\, 2 \,{t}^{6}+\, \,3\, {t}^{7}+ \,{t}^{8}
+ 2 \,{t}^{9}+3\,{t}^{10}  \, +\,{t}^{11} +\, 2\,{t}^{12} \, 
+3 \, t^{13}\, + \,3\,  t^{14}
+\, \ldots \nonumber
\end{eqnarray}
For  $\, \epsilon \,= 6 \,  $:
\begin{eqnarray}
\label{expanrealArnold1610}
{\cal A}_{6}(t) \, = \, \,t+{t}^{2}+2\,{t}^{3}+3\,{t}^{4}+\,{t}^{5}
+2\,{t}^{6}+3\,{t}^{7}+ 3 \,{t}^{8}
+ 2 \,{t}^{9}+3\,{t}^{10}  \, +3\,{t}^{11} +4\,{t}^{12} +3 \, t^{13}
+\, \ldots \nonumber
\end{eqnarray}
For  $\, \epsilon \,= 10 \,  $:
\begin{eqnarray}
\label{expanrealArnold161}
{\cal A}_{10}(t) \, = \,\, 
t+{t}^{2}+2\,{t}^{3}+3\,{t}^{4}+3\,{t}^{5}
+2\,{t}^{6}+3\,{t}^{7}+3\,{t}^{8}
+4\,{t}^{9}+5\,{t}^{10} \, +5\,{t}^{11} +4\,{t}^{12}+5\, {t}^{13}
+\, \ldots \nonumber
\end{eqnarray}
For  $\, \epsilon \,= 20 \,  $:
\begin{eqnarray}
\label{expanrealArnold162}
{\cal A}_{20}(t) \, =\,  \,t+{t}^{2}+2\,{t}^{3}+3\,{t}^{4}
+5\,{t}^{5}+4\,{t}^{6}+5\,{t}^{7}+3\,{t}^{8}+8\,{t}^{9} \, +11\,
t^{10} + \,7 \, t^{11} + \,10 \,  t^{12} + \, 21 \, t^{13}+ \, 
 \ldots \nonumber
\end{eqnarray}
For  $\, \epsilon \,= 30 \,  $:
\begin{eqnarray}
\label{expanrealArnold16330}
{\cal A}_{30}(t) \, = \, \,\,t+{t}^{2}+2\,{t}^{3}+3\,{t}^{4}
+5\,{t}^{5}+4\,{t}^{6}+5\,{t}^{7}+9\,{t}^{8}+8\,{t}^{9} \,+11\, t^{10}
+ 11\, t^{11}+14 \, t^{12} +25 \, t^{13}\, + \ldots \nonumber
\end{eqnarray}
For  $\, \epsilon \,= 40 \,  $ :
\begin{eqnarray}
\label{expanrealArnold163}
{\cal A}_{40}(t) \, =\,  \,\,t+{t}^{2}+2\,{t}^{3}+3\,{t}^{4}
+5\,{t}^{5}+4\,{t}^{6}+5\,{t}^{7}
+9\,{t}^{8}+10 \,{t}^{9}  \, +\,15 \, t^{10}+11 \, t^{11}+14 \, t^{12}
+ \, 29 \, t^{13}+ \ldots \nonumber
\end{eqnarray}
For  $\, \epsilon \,= 50 \,  $:
\begin{eqnarray}
\label{expanrealArnold16450}
{\cal A}_{50}(t) \, = \, \,\,t+{t}^{2}+2\,{t}^{3}+3\,{t}^{4}
+5\,{t}^{5}+4\,{t}^{6}+7 \,{t}^{7}
+9 \,{t}^{8}
+10 \,{t}^{9} \, +\, 15 \, t^{10}+\, 17 \, t^{11} +22 \, t^{12} + 37\, t^{13}
\, + \ldots \nonumber
\end{eqnarray}

For  $\, \epsilon \,= 100 \,  $:
\begin{eqnarray}
\label{expanrealArnold164}
{\cal A}_{100}(t) \, = \,\, \,\,t+{t}^{2}+2\,{t}^{3}+3\,{t}^{4}
+5\,{t}^{5}+8\,{t}^{6}+7 \,{t}^{7}
+9 \,{t}^{8}
+16 \,{t}^{9} \, +\, 19 \, t^{10}+\, 29 \, t^{11} + 36 \, t^{12} + 51 \, t^{13}
\, + \ldots \nonumber
\end{eqnarray}

\end{document}